\documentclass[apj]{emulateapj}
\usepackage{amsmath}
\usepackage{graphicx}
\usepackage{natbib}
\usepackage{epsfig}
\usepackage{multirow}
\usepackage{enumitem}

\usepackage{url}
\usepackage{indentfirst}
\usepackage{color}
\usepackage[normalem]{ulem}
\usepackage[left]{lineno}

\newcommand{\psr}{NGC\,7793 P13}

\shorttitle{Timing and spectral properties of NGC\,7793\,P13}
\shortauthors{Lin et al.}

\begin{document}

\title{Investigation of the timing and spectral properties of an Ultra-luminous X-ray pulsar NGC 7793 P13}     

\author{Lupin Chun-Che Lin${}^{1,4}$,
         Chin-Ping Hu${}^2$,  
         Jumpei Takata${}^3$,
         Kwan-Lok Li${}^4$,
         C. Y. Hui${}^5$, and
         A. K. H. Kong${}^6$
         }
 \affil{${}^1$
        Department of Physics, UNIST, 
        Ulsan 44919, Korea}
\affil{${}^2$
       Department of Physics, 
       National Changhua University of Education, 
       Changhua 50007, Taiwan}
\affil{${}^3$
       School of Physics, 
       Huazhong University of Science and Technology, 
       Wuhan 430074, China}      
\affil{${}^4$
       Department of Physics,
       National Cheng Kung University, 
       Tainan 701401, Taiwan;
       lupin@phys.ncku.edu.tw}               
 \affil{${}^5$
       Department of Astronomy and Space Science, 
       Chungnam National University, 
       Daejeon 305-764, Korea}        
\affil{${}^6$
	    Institute of Astronomy, 
	    National Tsing Hua University, Hsinchu 30013, 
	    Taiwan}   

\begin{abstract}
We perform both timing and spectral analyses using the archival X-ray data taken with \emph{Swift}, \emph{XMM-Newton}, \emph{NICER}, and \emph{NuSTAR} from 2016 to 2020 to study an ultra-luminous pulsar, NGC 7793 P13, that showed a long period of super-Eddington accretion.
We use the Rayleigh test to investigate the pulsation at different epochs, and confirm the variation of the pulse profile with the finite Gaussian mixture modelling and two-sample Kuiper test. 
Taking into account the periodic variation of the spin periods caused by the orbital Doppler effect, we further determine an orbital period of $\sim$65\,d and show that no significant correlation can be detected between the orbital phase and the pulsed fraction.
The pulsed spectrum of \psr\ in 0.5--20\,keV can be simply described using a power-law with a high energy exponential cutoff, while the broad-band phase-averaged spectrum of the same energy range requires two additional components to account for the contribution of a thermal accretion disk and the Comptonization photons scattered into the hard X-rays. 
We find that \psr\ stayed in the hard ultra-luminous state and the pulsed spectrum was relatively soft when the source was faint in the end of 2019.
Moreover, an absorption feature close to 1.3\,keV is marginally detected from the pulsed spectra and it is possibly associated with a cyclotron resonant scattering feature.

\end{abstract}


\section{Introduction}
\label{sec:intro}
Ultra-luminous X-ray sources (ULXs; \citealt{FS2011,KFR2017,Atapin2018}) are non-nuclear X-ray point-like sources with the X-ray emission exceeding the Eddington luminosity of a 10\,$M_{\odot}$ black hole ($L_x \gtrsim 10^{39}$\,ergs\,s$^{-1}$, and therefore most of these compact objects were usually suggested to host heavier black holes (e.g., intermediate-mass black holes) with isotropic radiation.
Nevertheless, the substantial beaming correlated with a geometrically thin accretion flow \citep{SM2003,Eksi2015,KMOO2016} for a source to radiate at an Eddington or a super-Eddington rate \citep{Liu2013,Bachetti2014} challenges the traditional model. Moreover, a neutron star/pulsar can also be a ULX.
The speculation for accreting neutron stars in ULXs was confirmed by the detection of coherent pulsations from X-ray point sources in nearby galaxies \citep{Bachetti2014}, and these sources are classified as ultra-luminous X-ray pulsars (ULPs).

Until early 2020, six known extragalactic pulsating ULXs have been confirmed to have a neutron star accretor; including M82 X-2 \citep{Bachetti2014}, NGC 5907 ULX1 \citep{Israel2017}, NGC 7793 P13 \citep{Furst2016,P13Israel2017}, NGC 300 ULX1 (SN2010da; \citealt{Carpano2018}), M51 ULX7 \citep{Rodriguez2020} and RX J0209.6$-$7427 \citep{Chandra2020}. 
Among these ULPs, a B9IA giant star was clearly identified for \psr\ \citep{Motch2011}, and this system demonstrated for the first time that super-Eddington accretion takes place in at least one ULX with an accretor's mass of $< 15$\,$M_{\odot}$.

Based on the  X-ray--UV--optical spectrophotometric monitoring program from 2009 to 2013, \citet{Motch2014} further detected a photometric period of $\sim$64 days that is associated with the radial velocity of He\,{\normalsize{II}} and the orbital period of the system.
The photometric maxima of the optical and UV bands reveal a phase jitter of up to 0.09, implying a superorbital modulation of 5--8.8\,yr.
A longer periodicity of 65.05\,d was obtained by a re-investigation of the photometric modulation in the X-ray band using the \emph{Neil Gehrels Swift Observatory} (\emph{Swift})/X-Ray Telescope (XRT) data \citep{Hu2017}.
The difference between the periods measured in the optical/UV and X-ray bands can be due to the beat frequency with a superorbital period caused by the precession of the accretion disk (or the funnel-shaped wind), and \citet{Hu2017} constrained the superorbital modulation within 2700--4700\,d.
The orbital ephemeris can be determined by the variation of the pulse/spin period as a function of orbital phase to account for the Doppler effect, and an orbital period of $\sim$63.9\,d with an eccentricity of $\leq$ 0.15 was therefore inferred through the variation of spin periods detected from {\it XMM-Newton} and the \emph{Nuclear Spectroscopic Telescope Array} ({\it NuSTAR}) observations \citep{Furst2018}.  

In comparison to a likely orbital period detected for NGC 5907 ULX-1 ($\sim$5\,d; \citealt{Israel2017}) and M51 ULX-7 ($\sim$2\,d; \citealt{RCastillo2020} and \citealt{HUE2021}) and a known period detected for M82 X-2 ($\sim$2.5\,d; \citealt{Bachetti2014}), we cannot reject the possibility that the periodicity of a few months yielded from the optical/UV or X-ray data of \psr\ is just another superorbital modulation, similar to the 78\,d \citep{Walton2016}, 38.5\,d \citep{Brightman2020,VLKB2020} and 55--62\,d \citep{Qiu2015,Kong2016} quasi-periodic signals detected for NGC 5907 ULX-1, M51 ULX-7 and M82 X-2, respectively.
For such a scenario, the observed phase jitter detected in the optical band can also represent an indication of a semi-periodic behavior, and the explanation of a long orbital period is not favored \citep{Furst2016}. 
The detection of the semi-periodic behavior can be illustrated by the Lense-Thirring precession of the accretion flow \citep{Middleton2015}, and it might also provide constraints on the equation of state of a neutron star. 

\citet{Walton2018} presented a detailed X-ray spectral analysis for \psr\ using the data before mid-2016. 
Two non-pulsed thermal blackbody components with temperatures of $\sim$0.5 and 1.5\,keV, and an additional continuum component that extends to the hard X-ray band ($\gtrsim 10$\,keV) associated with the pulsed emission from the accretion column are required to describe the broadband spectral behavior.
A much harder power-law ($\Gamma \lesssim 0$) with a lower high energy cutoff ($E_{\rm cut}\sim 4$\,keV) in comparison to other ULPs can provide a good fit to the pulsed spectrum.
In this study, we concentrate on the timing and spectral properties inferred from the X-ray data sets observed with higher cadence in 2016--2020.
The observed flux of \psr\ was in the high state in mid-2016, and it has significantly dropped below the detection threshold of a {\emph Swift}/XRT monitoring campaign since the end of 2019.
We describe the archival data obtained from different X-ray missions in Section~\ref{sec:observations}.
We searched for the pulsation, determined the spin properties and compared the pulsed structure at different epochs in Section~\ref{ssec:Panalysis}.
The X-ray modulation and orbital properties are investigated and determined in Section~\ref{ssec:Oanalysis}, while we checked the broad-band and pulsed spectral behaviors in Section~\ref{ssec:Sanalysis}.
According to all the results mentioned in Section~\ref{sec:results}, we followed the known scenarios to discuss the major observed timing and spectral features in our investigation in Section~\ref{sec:Discussion}, and finally we provide a summary in Section~\ref{sec:Summary}.

\section{High-energy Observations}
\label{sec:observations}

To reveal the X-ray emission nature of \psr, we collected archival data observed with \emph{Swift}, \emph{XMM-Newton}, \emph{the Neutron star Interior Composition Explorer} (\emph{NICER}), and \emph{NuSTAR} in our investigation.
We used \emph{Swift} data to examine the 65-day X-ray modulation for \psr. 
The archival data obtained from other three missions have much more X-ray photons, enabling a detailed study of the pulsed emission and spectral behavior of our target.
Except for \emph{NICER} observations, we considered the pulsar position at R.A.=$23^{\rm h}57^{\rm m}59\fs{9}$, decl.=$-32^{\circ}37'26\farcs6$ (J2000) according to the  \emph{Chandra} counterpart, CXOU J235750.9$-$323726 \citep{Pannuti2011}, to extract source photons.
Photon arrival times applied in the periodicity search were corrected to the barycentric dynamical time (TDB) using the JPL DE405 solar system ephemeris at the aforementioned source position. 

\subsection{{\sl Swift}}
\label{ssec:Swift}
We collected all the data sets observed by the XRT on board \emph{Swift}, and these observations were operated under the photon counting (PC) mode with a time resolution of 2.5\,s.
Due to the limit of the time resolution and a short duration of these PC mode observations, we did not check the spin period using these data.
Nevertheless, the long term monitoring of \psr\ since 2010 Aug. 16 (i.e., MJD~55424) provides us an opportunity to survey the source emission at high or faint stages and to examine the modulation caused by the orbital motion. 
All the \emph{Swift} data and the related products were obtained from the {\emph{Swift} website\footnote{https://www.swift.ac.uk/user\_objects/}} \citep{Evans2007,Evans2009}, and all the processes were performed using the HEASoft package (v.6.28; \citealt{HEAsoft2014}).
We obtained the light curve with each bin corresponding to a single observation.  
In the above data reduction, only photon energies in the range of 0.3--10 keV with grades 0--12 were included to generate the light curve.

\subsection{{\sl XMM-Newton}}
\label{ssec:XMM}
\emph{XMM-Newton} observed \psr\ twelve times from 2016 to 2020.
The EPIC cameras on board \emph{XMM-Newton} observed \psr\ with the thin filter and the full frame mode, which has a temporal resolution of 2.6\,s and 73.4\,ms for MOS (Metal Oxide Semi-conductor) and pn cameras, respectively.
The \emph{XMM-Newton} data provide an opportunity to investigate the spectral behavior in the soft X-ray band (0.2--10\, keV) owing to the large effective area and long exposure. 
Those observed by the pn camera have sufficient temporal solution to examine the spin pulsation of our target.
We list all the \emph{XMM-Newton} data sets for pulsation studies between 2016 May 20 and 2019 November 22 in Table~\ref{tsolution}.
To reduce the source events for subsequent timing and spectral analyses, we used the \emph{XMM-Newton} Science Analysis Software (XMMSAS; version 16.1.0; \citealt{SAS2004}). We kept single- to double-pixel events (PATTERN = 0--4) for the pn camera and single- to quadruple-pixel events (PATTERN = 0--12) for the two MOS cameras.
We also set ``FLAG==0'' in our data selection expression to filter out artifacts from the calibrated and concatenated data sets and then removed photons collected in X-ray background flares by comparing count rates accumulated in a short time bin. 
The source events were extracted in the 0.15--12\,keV band from a 25\arcsec\ or 30\arcsec\ radius region centered at the \emph{Chandra} position mentioned in Section~\ref {sec:observations}.
The size of the selection region depends on the contour levels of the image, and we ensure the region contains $>$80\% encircled energy.

\begin{table*}
\caption{\footnotesize{Measurements of pulsation and flux.}}\label{tsolution} 
\centering
\begin{tabular}{llllcllclll} 
\hline
\hline
Mission & Time & Date (MJD) & ObsID & EXP (ks) & Frequency (Hz) & $\dot{F}$ (10$^{-10}$\,s$^{-2}$) & $Z_1^2$/trials & phase$^{b}$ & PF$^{c}$ & Flux$^{e}$
\\
\hline
NuSTAR & 2016-05-20 & 57528.18 & \dataset[https://heasarc.gsfc.nasa.gov/FTP/nustar/data/obs/02/8/80201010002]{80201010002} & 106 & 2.398361(6) &  0.2$\pm 0.6$ & 795.4/77k & 0.97(0) & $32\pm 2$\% & $36.5^{+0.7}_{-0.6}$ 
\\
XMM & 2016-05-20 & 57528.58 & \dataset[http://nxsa.esac.esa.int/nxsa-web/\#obsid=0781800101]{0781800101} & 47 & 2.39836(3) & $0.9\pm 13.1$ & 326.2/1k & 0.98(0) & $16\pm 2$\% & $31.8\pm 0.8$ 
\\
XMM & 2017-05-13 & 57886.17 & \dataset[http://nxsa.esac.esa.int/nxsa-web/\#obsid=0804670201]{0804670201} & 25 & $\cdots^{a}$ & $\cdots^{a}$ & $\cdots^{a}$/0.2k & 0.49(1) & $< 10$\%$^{d}$ & $14.0\pm 0.9$
\\
NuSTAR & 2017-05-19 & 57892.71 & \dataset[https://heasarc.gsfc.nasa.gov/FTP/nustar/data/obs/03/3/30302005002]{30302005002} & 83 & 2.404565(6) &  $8.1\pm 0.7$ & 82.5/44k & 0.60(1) & $17\pm 5$\% & $16.2\pm 0.5$
\\
XMM & 2017-05-20 & 57893.66 & \dataset[http://nxsa.esac.esa.int/nxsa-web/\#obsid=0804670301]{0804670301} & 55 & 2.40463(2) & $8.3\pm 7.0$  & 79.8/1.6k & 0.61(1) & $12\pm 2$\% & $16.9\pm 0.5$
\\
XMM & 2017-05-31 & 57904.90 & \dataset[http://nxsa.esac.esa.int/nxsa-web/\#obsid=0804670401]{0804670401} & 31 & 2.40543(4) & $17.3\pm 24.7$ & 49.1/0.2k & 0.78(1)  & $10\pm 2$\% & $28.0\pm 0.8$
\\
XMM & 2017-06-12 & 57916.10 & \dataset[http://nxsa.esac.esa.int/nxsa-web/\#obsid=0804670501]{0804670501} & 32 & 2.40587(4) & $-5.8\pm 26.2$ & 44.9/0.2k & 0.96(1) & $10\pm 2$\% & $38.6^{+1.0}_{-1.1}$ 
\\
XMM & 2017-06-20 & 57924.11 & \dataset[http://nxsa.esac.esa.int/nxsa-web/\#obsid=0804670601]{0804670601}  & 30 & 2.40581(4) & $7.4\pm 30.7$ &  45.7/0.2k & 0.08(1) & $9\pm 2$\% & $35.9^{+1.0}_{-0.9}$
\\ 
NuSTAR & 2017-06-29 & 57933.93 & \dataset[https://heasarc.gsfc.nasa.gov/FTP/nustar/data/obs/03/3/30302015002]{30302015002} & 76 &  2.405553(8) & $-3.4\pm 1.1$ & 108.6/31k & 0.23(1) & $20\pm 3$\% & $33.3\pm 0.8$
\\
NuSTAR & 2017-07-08 & 57942.93 & \dataset[https://heasarc.gsfc.nasa.gov/FTP/nustar/data/obs/03/3/30302015004]{30302015004} & 75 & 2.405340(7) & $-0.3\pm 0.8$ & 60.7/36k & 0.37(1) & $17\pm 3$\% & $26.7\pm 0.7$
\\
NICER & 2017-10-31 & 58057.01 & \dataset[https://heasarc.gsfc.nasa.gov/FTP/nicer/data/obs/2017_10/1050420102]
{1050420102} & 11 & 2.408503(8) & $-3.5\pm 1.3$ & 87.1/6.2k & 0.13(2) & $13\pm 2$\% & $39.7^{+3.8}_{-3.6}$
\\
NuSTAR & 2017-10-31 & 58057.58 & \dataset[https://heasarc.gsfc.nasa.gov/FTP/nustar/data/obs/03/9/90301326002]{90301326002} & 52 & 2.40849(1) & $-2.5\pm 2.2$ & 337.0/11k & 0.14(2) & $29\pm 3$\% & $44.1\pm 1.1$
\\
NuSTAR & 2017-11-25 & 58082.95 & \dataset[https://heasarc.gsfc.nasa.gov/FTP/nustar/data/obs/03/3/30302005004]{30302005004} & 77 & 2.408391(8) & $6.9\pm 1.0$ & 202.6/32k & 0.53(2) & $24\pm 3$\% &  $25.6^{+0.6}_{-0.7}$
\\
XMM & 2017-11-25 & 58083.00 & \dataset[http://nxsa.esac.esa.int/nxsa-web/\#obsid=0804670701]{0804670701} & 50 & 2.40840(2) & $6.1\pm 9.5$ & 274.5/1.2k & 0.53(2) & $19\pm 2$\% & $25.2^{+0.6}_{-0.7}$ 
\\
XMM & 2018-11-27 & 58449.79 & \dataset[http://nxsa.esac.esa.int/nxsa-web/\#obsid=0823410301]{0823410301} & 25 & 2.41767(5) & $-3.9\pm 37.6$ & 587.4/0.1k & 0.19(3) & $29\pm 3$\% & $19.8^{+0.7}_{-0.8}$ 
\\
XMM & 2018-12-27 & 58479.61 & \dataset[http://nxsa.esac.esa.int/nxsa-web/\#obsid=0823410401]{0823410401} & 25 & 2.41792(5) & $9.6\pm 37.8$ & 474.9/0.1k & 0.65(3) & $42\pm 3$\% & $12.5^{+0.7}_{-0.6}$ 
\\
XMM & 2019-05-16 & 58620.00 & \dataset[http://nxsa.esac.esa.int/nxsa-web/\#obsid=0840990101]{0840990101} & 43 & 2.42039(3) & $6.9\pm 18.3$ &  553.6/0.4k & 0.82(4) & $33\pm 3$\% & $9.1^{+0.5}_{-0.4}$ 
\\
NuSTAR & 2019-11-18 & 58805.69 & \dataset[https://heasarc.gsfc.nasa.gov/FTP/nustar/data/obs/04/5/50401003002]{50401003002} & 43 & 2.42346(1) & $8.1\pm 3.6$ & 142.1/5.9k & 0.68(4) & $55\pm 8$\% &  $3.4\pm 0.4$
\\
XMM & 2019-11-22 & 58809.40 & \dataset[http://nxsa.esac.esa.int/nxsa-web/\#obsid=0853981001]{0853981001} & 50 & 2.42374(3) & $9.3\pm 13.9$ & 441.5/0.9k & 0.74(4) & $46\pm 4$\% & $4.4^{+0.2}_{-0.3}$
\\
NuSTAR & 2019-12-13 & 58830.95 & \dataset[https://heasarc.gsfc.nasa.gov/FTP/nustar/data/obs/05/3/30502019002]{30502019002} & 77 & 2.424537(7) & $-1.6\pm 1.0$ & 188.3/11k  & 0.07(4) & $50\pm 7$\% & $3.4\pm 0.3$ 
\\
NuSTAR & 2020-01-08 & 58856.54 & \dataset[https://heasarc.gsfc.nasa.gov/FTP/nustar/data/obs/05/3/30502019004]{30502019004} & 52 & 2.42378(1) & $-0.3\pm 2.2$ & 185.1/12k & 0.46(4) &  $49\pm 7$\% & $3.9^{+0.3}_{-0.4}$
\\
NuSTAR & 2020-08-22 & 59083.22 & \dataset[https://heasarc.gsfc.nasa.gov/FTP/nustar/data/obs/06/9/90601327002]{90601327002} & 51 & 2.42762(1) & $5.5\pm 1.7$ & 30.4/11k & 0.95(5) &  $31\pm 9$\% & $3.5\pm 0.4$
\\
\hline
\hline
\end{tabular}
\begin{flushleft}
{\footnotesize
$^{a}$: The significance of pulsation is too low to be certainly confirmed.\\
$^{b}$: Orbital phase inferred from the elliptical orbital period of $64.84\pm 0.13$\,d (Table~\ref{orbitpara}). Uncertainties of phase are measured by the propagation of the uncertainty of the orbital period, and time zero is determined at MJD 57530. The uncertainty is labelled as 0 if it is smaller than 0.01.\\ 
$^{c}$: Pulsed fraction measured with $\frac { \rm{max}(N_{ i })-\rm{min}(N_{ i }) }{ \rm{max}(N_{ i })+\rm{min}(N_{ i }) }$, where $N_{ i }$ is the number of photons in a 25-bin pulse profile.\\ 
$^{d}$: 3$\sigma$ pulsed fraction upper limit for the extracted region assessed by the 0.5 duty cycle through eq. (26) in \citet{Jager94}.\\
$^{e}$: Unabsorbed flux in $10^{-13}$\,ergs\,s$^{-1}$\,cm$^{-2}$ determined in the 3--10\,keV energy band, and the uncertainty corresponds to 90\% confidence level.\\
}
\end{flushleft}
\end{table*}

We performed barycentric correction using the XMMSAS task \texttt{barycen}. 
For spectral analysis, we determined the background in a nearby source-free region with the same size as the source region and regrouped the channels to have $>$25 photons per channel for each observation to ensure $\chi^2$ statistics.
We generated the response matrices and ancillary response files with the XMMSAS tasks \texttt{rmfgen} and \texttt{arfgen}.
To generate the phase-resolved spectrum, we created GTI files according to the timing solution in Table~\ref{tsolution} and extracted the spectra using the XMMSAS tasks \texttt{evselect}. 

\subsection{{\sl NICER}}
\label{ssec:NICER}
\emph{NICER} observed \psr\ with the X-ray timing instrument (XTI) thrice from 2017 October 30 to November 1 during the high flux state.
We notice that this instrument has a sensitivity four times better than \emph{XMM-Newton} in the soft X-ray band (0.2--12\,keV) with a very precise timing resolution ($< 300$\,ns) although the observation obtained from the XTI does not have an imaging capability \citep{Arzoumanian2014,Okajima2016}.
In this paper, we concentrated on the analysis of the data set with the longest exposures ($\sim$11\,ks) obtained on 2017 Oct 31 because the accumulated exposure of two other data sets is less than 4\,ks which cannot yield a significant pulse detection.
In Table~\ref{tsolution}, we also present the information of the \emph{NICER} data used in our investigation.
We used a cleaned event list processed by the \texttt{nicerl2} pipeline script.
In the timing analysis, we only extracted photons within the energy range of 0.25--12\,keV since a significant noise peak can be resolved below this energy range to contaminate the pulsed detection.
We kept the photons collected from all 52 \emph{NICER} focal plane modules (FPMs) because no significant detector noise was found in a specific FPM.    
All the photon arrival times are corrected to the solar system barycenter using the \texttt{barycorr} task.
 
For the phase-averaged spectral analysis, we used the latest redistribution matrix and ancillary response files (i.e., nixtiref20170601v002.rmf and nixtiaveonaxis20170601v004.arf) generated by the \emph{NICER} team to account for the response from the XTI.
We compute the background using the version 6 of the \texttt{nibackgen3C50} tool \citep{Remillard2021} by taking a target observation events and pulling out the proxy data to build a predicted background spectrum.
In the preparation for the phase-resolved spectrum, we created GTI files for on- and off- pulse phase intervals with respect to the full-width half-maximum (FWHM) of the major pulsation determined from the folded light curve. 
We used the same data reduction procedures employed in the \emph{XMM-Newton} data (see \S2.2) to extract the phase-resolved spectra and group the spectra for spectral analysis. 
We performed all the \emph{NICER} data reduction using HEASoft (v.6.22; \citealt{HEAsoft2014}).

\subsection{{\sl NuSTAR}}
\label{ssec:NuSTAR}
\emph{NuSTAR} observed \psr\ with the onboard Focal Plane Modules A and B (FPMA/B) and provided the data with a temporal precision of $\sim$100\,$\mu$s \citep{Bachetti2021}.
Except for a very short exposure (0.4\,ks) on 2019 December 13, all other \emph{NuSTAR} observations since 2016 May have exposures of more than 50\,ks, providing enough photons to investigate the spin period.
In addition, \emph{NuSTAR} observations cover an effective energy range of 3--79\,keV \citep{Harrison2013}, which allow us to investigate the timing and spectral behavior of the harder X-ray band in comparison to the \emph{XMM-Newton} and \emph{NICER} data.
Here we also list all the \emph{NuSTAR} observations used in this investigation in Table~\ref{tsolution}.

For the data reduction of \emph{NuSTAR} observations, we used the HEASoft package together with NuSTARDAS v1.8.0 and the \emph{NuSTAR} calibration database (CALDB version 20210104). 
We adopted a source region of 50\arcsec\ radius, and the source events were constrained within a pulse-invariant channel of 35--1909, which corresponds to an energy range of 3--79\,keV.
We generated the energy spectra of the source and background, and related response matrices using the \texttt{nuproducts} tasks with the default setting. The background was extracted with an annular region centered at our target's position with an inner and outer radius of 50 pixels ($\sim$123\arcsec) and 80 pixels ($\sim$197\arcsec), respectively.
We note that the `BACKSCAL' keyword in the fits extension of the background spectrum in our analysis is also updated for the exposure variations within the extraction region.
Similar to the \emph{XMM-Newton} observations, we also regrouped the channels to have at least 25 or 50 photons per channel for each FPMA/FPMB datum in the spectral analysis.
In timing analysis, we corrected the photon arrival times to the solar system barycenter using the \texttt{barycorr} task.
To obtain the pulsed spectrum, we also generated GTI files for on- and off- pulse phase intervals following the similar procedure in reducing \emph{XMM-Newton} and \emph{NICER} observations.  
The above processes can be accomplished by employing the GTI files in the \texttt{nuproducts} task with the `usrgtifile' parameter.

\section{Results}
\label{sec:results}
We investigated the timing and spectral properties of \psr\ at different epochs using the X-ray observations as described in Section~\ref{sec:observations}. In the following subsections, we will present the detailed data analysis and results.

\subsection{{\sl Pulsed detection and structure}}
\label{ssec:Panalysis}

Different from the $\chi^2$-statistic methods applied in \citet{Furst2016,Furst2018,Furst2021} to inspect the spin periodicity of \psr, we obtained the pulsation via the Rayleigh test \citep{Mardia72,Gibson82}. 
This method can provide a better resolution and concentration for pulse detection.
We only searched for the pulsed detection in a 2-dimensional space within a small interval of frequency and frequency derivative extrapolated from the known timing solution provided by \citet{Furst2018}.
Such a small 2-d interval adopted in our search roughly covers an interval of $\sim 10^{-3}$\,Hz in frequency and $10^{-8}$\,Hz\,s$^{-1}$ in frequency derivative.
The inferred independent trials listed in Table~\ref{tsolution} is referred to the different Fourier width in each search.  
We reported the timing solution at different epochs by fitting the powers of the spin frequency and its derivative with a Gaussian function. 
The uncertainties of the timing parameters were estimated from the FWHM of the Gaussian.
Compared with \citet{Furst2018}, we did not obtain any signal using the \emph{XMM-Newton} data on 2017-05-13. 
We detected a marginal pulsation from the \emph{NuSTAR} data observed on 2020-08-22.
The chance probability to yield a signal with a Rayleigh-power ($Z_1^2$) over 30.4 among 11000 trials corresponds to 2.7$\times 10^{-3}$, meaning that the significance level of our detection on 2020-08-22 is equivalent to 3$\sigma$.

Except for the \emph{XMM-Newton} observation of 2017-05-13, we detect the pulsation from all the other data sets more significant than the detection on 2020-08-22.
We generated the folded light curve using 25 bins and estimated the pulsed fraction (PF) as defined in note c of Table~\ref{tsolution}.
We list the related results in the PF column of Table~\ref{tsolution}. 
The $Z_1^2$ power and the PF detected from the \emph{NICER} data are all significantly lower than those determined from the \emph{NuSTAR} observation in the same energy band and the similar epoch because \emph{NICER} observations have no imaging capability to precisely extract source photons.
PFs shown in Table~\ref{tsolution} significantly depend on the number of bins to fold the profile and we therefore also consider to estimate the PFs based on the Fourier decomposition with harmonics numbers $k < 6$ to reduce the binning effect \citep{DKG2009,HNH2019}.
Results of the new definition in PF yield about an half of those values listed in Table~\ref{tsolution} and remain a similar distribution in time.

\psr\ showed a single-pulsed, nearly sinusoidal pulse profile in most of the X-ray data sets.
However, the pulse profiles in mid-2017 seem to be less sinusoidal and show a narrower peak in both \emph{XMM-Newton} and \emph{NuSTAR} observations, and this is consistent with the result of \citet{Furst2018}.
In order to statistically investigate the change of the pulsed structures at different epochs, we used the finite Gaussian mixture modelling (GMM; details in Appendix~\ref{app:GMM}) to determine the number of Gaussian components existed in the pulse profiles. 
Most of the profiles can be depicted with a single peak of one Gaussian, but the profile was composed of two Gaussians in the \emph{XMM-Newton} data from late 2018 to mid-2019, which potentially represents a much broader pulsation shown in the profile.
Table~\ref{FWHM} summarizes the full-width at half-maximum (FWHM) of Gaussians determined via GMM method at different epochs.
We find that the pulse profile shows a narrower peak in 2017 May, and the width of the peak became broader from mid-2017 to mid-2019.  
The pulsed emission in the soft X-ray band (i.e., $\lesssim 10$\, keV) can be described with a single Gaussian again at the end of 2019, and it became weak (or disappeared) in mid-2020.

\begin{table*}
\caption{\footnotesize{FWHM of the Gaussian fitted to the pulsation.}}\label{FWHM} 
\centering
\begin{tabular}{ccccccccccc} 
\hline
\hline
Mission & \scriptsize{2016-05-20} & \scriptsize{2017-05-19} & \scriptsize{2017-06-29} & \scriptsize{2017-07-08} & \scriptsize{2017-10-31} & \scriptsize{2017-11-25} & \scriptsize{2019-11-18} & \scriptsize{2019-12-13} & \scriptsize{2020-01-08} & \scriptsize{2020-08-22}
\\
\hline
NuSTAR & 0.37(3) & 0.30(3) & 0.33(2) & 0.35(2) & 0.37(1) & 0.35(1) & 0.35(2) & 0.36(2) & 0.34(2) & 0.37(2)
\\
\hline
  & \scriptsize{2016-05-20} & \scriptsize{2017-05-20} & \scriptsize{2017-05-31} & \scriptsize{2017-06-12} & \scriptsize{2017-06-20} & \scriptsize{2017-11-25} & \scriptsize{2018-11-27} & \scriptsize{2018-12-27} & \scriptsize{2019-05-16} & \scriptsize{2019-11-22} 
\\
\hline
XMM & 0.36(2) & 0.31(2) & 0.32(2) & 0.33(2) & 0.33(2) & 0.346(7) & \scriptsize{0.26(3), 0.28(1)} & \scriptsize{0.32(4), 0.21(4)} & \scriptsize{0.27(2), 0.27(1)} & 0.35(1)
\\
\hline
 & \scriptsize{2017-10-31} &  &  &  &  &  &  &  &  &  
\\
\hline
NICER & 0.34(2) &  &  &  &  &  &  &  &  & 
\\
\hline
\hline
\end{tabular}
\begin{flushleft}
{\footnotesize
Notes. The numbers in parentheses denote errors in the last digit.}
\end{flushleft}
\end{table*}

\begin{table*}
\caption{\footnotesize{Comparison of two profiles with Kuiper test.}}\label{KS-test} 
\centering
\begin{tabular}{cccccccccc} 
\hline
\hline
\multirow{3}{*}{NuSTAR} & 2016-05-20 & 2017-05-19 & 2017-06-29 & 2017-07-08 & 2017-10-31 & 2017-11-25 & 2019-11-18 & 2019-12-13 & 2020-01-08 
\\
 & vs & vs & vs & vs & vs & vs & vs & vs & vs 
\\
 & 2017-05-19 & 2017-06-29 & 2017-07-08 & 2017-10-31 & 2017-11-25 & 2019-11-18 & 2019-12-13 &  2020-01-08 & 2020-08-22     
\\
\hline
Prob. & 3.2$\times 10^{-9}$ & 0.52 & 0.37 & 2.8$\times 10^{-8}$ & 0.05 & 1.9$\times 10^{-8}$ & 0.60 & 0.95 & 3.0$\times 10^{-6}$  
\\
\hline
\multirow{3}{*}{NuSTAR} & 2016-05-20 & 2016-05-20 & 2017-05-19 & &  &  &  &  &  
\\
 & vs & vs & vs &  &  &  &  & &  
\\
 & 2017-10-31 & 2019-11-18 & 2020-08-22 &  &  &  &  &   &  
\\
\hline
Prob. & 0.011 & 2.5$\times 10^{-4}$ & 0.52 &  &  &  &  &  &   
\\
\hline
\hline
\multirow{3}{*}{XMM} & 2016-05-20 & 2017-05-20 & 2017-05-31 & 2017-06-12 & 2017-06-20 & 2017-11-25 & 2018-11-27 & 2018-12-27 & 2019-05-16 
\\
 & vs & vs & vs & vs & vs & vs & vs & vs & vs  
\\
 & 2017-05-20 & 2017-05-31 & 2017-06-12 & 2017-06-20 & 2017-11-25 & 2018-11-27 & 2018-12-27 &  2019-05-16 & 2019-11-22   
\\
\hline
Prob. & 5.6$\times 10^{-6}$ & 0.68 & 0.88 & 0.99 & 1.4$\times 10^{-8}$ & 7.9$\times 10^{-22}$ & 0.19 & 0.83 & 1.8$\times 10^{-3}$ 
\\
\hline
\multirow{3}{*}{XMM} & 2016-05-20 & 2016-05-20 & 2016-05-20 & 2017-11-25 &  &  &  &  &  
\\
 & vs & vs & vs & vs &  &  &  & &  
\\
 & 2017-11-25 & 2018-11-27 & 2019-11-22 & 2019-11-22 &  &  &  &   &  
\\
\hline
Prob. & 0.23 & 1.8$\times 10^{-17}$ & 3.2$\times 10^{-27}$ & 2.2$\times 10^{-30}$   &  &  &  &  &   
\\
\hline
\hline
\end{tabular}
\end{table*}

We then performed a two-sample Kuiper test (referred to Appendix~\ref{app:Kuiper-test}) on the unbinned phase distributions to examine the significance of the variation between two pulse profiles. 
Corresponding results are summarized in Table~\ref{KS-test}. It is worth noting that we did not compare with the arrival phase information obtained from different missions since they cover different energy range.
Accompanying with the variation of the Gaussian width shown in Table~\ref{FWHM}, we can further confirm that the pulse structure is similar when the PF is low (i.e., $\lesssim 20$\% for \emph{NuSTAR} and  $\lesssim 10$\% for \emph{XMM-Newton} observations) or the pulsed detection is relatively weak (i.e., $\lesssim 100$ for $Z_1^2$).
Except for the pulse structure change occurred between the strong and weak pulsed detection/fraction, we also found that such a variation existed even when the pulsed fraction/detection remained in the strong stage.
For instance, though the differentiation of the the pulse profiles between 2016-05-20 and 2017-10-31 (for \emph{NuSTAR})/2017-11-25 (for \emph{XMM-Newton}) is less than 3$\sigma$, it is much more significant between 2017-11-25 and 2019-11-18 (for \emph{NuSTAR})/2019-11-22 (for \emph{XMM-Newton}). 
We note that the broadening of the main peak between the end of 2017 and of 2018 is indicated in Table~\ref{FWHM}, and the Kuiper test can further verify such a structural change in pulsation. 

\subsection{{\sl X-ray modulation and orbital period}}
\label{ssec:Oanalysis}

We applied the Lomb-Scargle method \citep{Lomb76,Scargle82} to examine the long-term X-ray modulation of the \emph{Swift} data mentioned in Section~\ref{ssec:Swift}.
The most significant signal shown in the periodogram is $P_{\rm x} = 65.6 \pm 0.5$\,days, while the uncertainty was determined by Monte Carlo simulations.
We note that the detected period is significant at more than 99\% confidence level determined from both white \citep{HB86} and red noise model \citep{SM2002} for the power spectrum of the original data.
Our latest detection for the X-ray modulation is between two previous detections (i.e., $65.05\pm 0.10$\,days in \citet{Hu2017} and $66.8\pm 0.4$ in \citet{Furst2018}) and is consistent with the recent report of $65.31\pm 0.15$\,days in \citet{Furst2021}. 

Detections of the aforementioned X-ray modulation for \psr\ are longer than all the periodic modulations ($P_{\rm opt} \lesssim 64$\,days) determined in the optical/UV band in the literatures \citep{Motch2014, Hu2017, Furst2018, Furst2021}. 
Since a longer superorbital modulation ($P_{\rm sup}$) of 5--13 years was also detected in \citet{Motch2014} and \citet{Hu2017}, $P_{\rm opt}$ can be explained by the beat period of $P_{\rm x}$ and $P_{\rm sup}$ if we treat $P_{\rm x}$ as the orbital period.
\citet{Motch2014} showed that the detection of $P_{\rm opt}$ is much more significant during the X-ray low state, suggesting that $P_{\rm x}$ and  $P_{\rm opt}$ are originated from different emission mechanisms.
In such a scenario, both $P_{\rm x}$ and $P_{\rm opt}$ can be related to the orbital modulation, but $P_{\rm opt}$ has a shorter period because of the phase jitter (e.g., Fig.~5(b) of \citealt{Hu2017}).
The X-ray modulation ($P_{\rm x}$) can be originated from a resonance \citep{WK91} between the Keplerian velocity and the orbital period if the emission is associated with the inner accretion disk. 
Therefore, it is not surprising that the detection of the X-ray modulation for \psr\ at different time intervals is not stable.  

A more precise method to estimate the orbital period $P_{\rm orb}$ can be obtained by assuming a model to count for the Doppler shift of a binary system, especially when we have detected the pulsation and determined the spin frequency at different epochs in Table~\ref{tsolution}. 
Following the similar concern proposed in \citet{Furst2018} and \citet{Furst2021}, we fixed the time/epoch zero ($T_0$) at MJD 57530.0 to perform a non-linear fit using the GNU scientific library to the distribution of the detected spin frequencies with an elliptical orbit of 7 free parameters including $P_{\rm orb}$, the projected semi-major axis ($a$\,sin$i$), the orbital eccentricity ($e$), epoch in MJD at which the mean orbital longitude of 90 degree ($T_{90}$), the argument of periapsis ($\omega$), a spin period ($P_{\rm spin}$) and a constant spin-up rate of the neutron star ($\dot{P}_{\rm spin}$). 
Since the pulsed detection on 2020-08-22 is very marginal, we did not include this data point in our fit.
Unfortunately, we did not obtain an acceptable fit since our timing errors determined in each pulsed detection are obviously smaller than those reported in \citet{Furst2021} due to a much better concentration of the 2-D detection contours presented with the Rayleigh test. 
It does not indicate the best-fit orbital parameters determined with such a method are not reliable, and we note that the true uncertainties of each detection might be underestimated without a good strategy to count for the systematic errors in our computation. 
Table~\ref{orbitpara} demonstrates the obtained orbital parameters. 
The small $e$ may suggest that a circular orbit is enough to describe the orbital motion, and we also find that the null hypothesis probability to improve the fit with an elliptical orbit is 0.59, which means that the effectiveness to consider the eccentricity to improve the fit is negligible.

The orbital period seems to be extended if we compare the results determined by the fits in \citet{Furst2018} and \citet{Furst2021}, and we also found that a longer orbital period can be assessed when we included more recent data points to fit the orbital motion.
We therefore also included a constant increasing rate of the orbital period in a circular orbit to fit the obtained spin detection as shown in Table~\ref{orbitpara}. 
Nevertheless, we note that such an additional parameter in the model cannot significantly improve the fit via a F-test, and the derived derivative of the orbital period in the best-fit is too small to cause any major effect in the observed time intervals.      

\begin{table}
\caption{\footnotesize{Best-fit orbital parameters obtained via the $\chi^2$ minimization method.}}\label{orbitpara} 
\centering
\begin{tabular}{l|ll|l} 
\hline
\hline
  & \multicolumn{2}{c}{Circular} &  \multicolumn{1}{|c}{Elliptical}
\\
Parameter & \multicolumn{2}{c}{orbit} & \multicolumn{1}{|c}{orbit}
\\
\hline
$P_{\rm spin}$\,(ms) & 417.075(8) & 417.1(7) & 417.08(1)   
\\
$\dot{P}_{\rm spin}$\,($10^{-11}$\,s\,s$^{-1}$) & -4.08(1) & -4(1) & -4.09(1)   
\\
$P_{\rm orb}$\,(d) & $64.86 \pm 0.09$ & $64.9 \pm 4.7$  & $64.84 \pm 0.13$ 
\\
$\dot{P}_{\rm orb}$\,(d\,d$^{-1}$) & $\cdots$ & 0.00005(4) & $\cdots$  
\\
$a$sin$i$\,(lt-s) & $276 \pm 17$ & $278 \pm 21$  & $273 \pm 20$ 
\\
T$_{90}$\,(MJD) &  56731(2) & 56730.4(9) & 56731(3)  
\\
$e$ & $\cdots$  & $\cdots$ & $0.03 \pm 0.03$   
\\
$\omega$ & $\cdots$ & $\cdots$  & $-151 \pm 71$  
\\
\hline
\end{tabular}
\begin{flushleft}
{\footnotesize
Note: The quoted uncertainties are in 1$\sigma$ confidence level..\\
}
\end{flushleft}
\end{table}

\begin{table*}
\caption{\footnotesize{Best-fitting spectral parameters of the broad-band continuum model (DISKBB+CUTOFFPL$\otimes$SIMPL) for \psr.}}\label{broadband-spectrum} 
\centering
\begin{tabular}{lccccccccccc} 
\hline
\hline
Parameters & $N_{\rm H; int}$ & $kT_{\rm in; DBB}$ & Norm$_{\rm DBB}$ &$^{a}$Flux$_{\rm DBB}$ &  $\Gamma_{\rm CPL}$ & $E_{\rm cut; CPL}$ & Norm$_{\rm CPL}$ & $\Gamma_{\rm SIMPL}$ & f$_{\rm scat}$ & $\chi_{\nu}^2$/dof 
\\
  & (10$^{20}$\,cm$^{-2}$) & (keV) &  & (\%) &  & (keV) &  &  &  (\%) &  &  
\\
\hline
2016 May 20-22 & $5.3_{-1.1}^{+1.2}$ & $0.41\pm 0.05$ & $1.5_{-0.5}^{+0.9}$ & 15.4 & $-0.54_{-0.38}^{+0.30}$ & $2.4_{-0.4}^{+0.6}$ & $2.3_{-0.3}^{+0.5} \times 10^{-4}$ & $3.5_{-2.2}^{+1.1}$ & $> 29.8$ & 0.99/1175
\\
2017 May 19-21 & $5.8_{-0.9}^{+1.4}$ & $0.39\pm 0.04$ & $1.3_{-0.3}^{+0.8}$ & 19.3 &  $-0.51_{-0.31}^{+0.10}$ & $2.4\pm 0.4$ & $1.7_{-0.5}^{+0.1} \times 10^{-4}$ &  $1.1_{-0.1}^{+1.1}$ & $> 34.2$ & 1.04/746
\\
2017 Oct 31-Nov 1 & $0.5_{-0.5}^{+3.3}$ & $0.35_{-0.07}^{+0.06}$ & $3.2_{-0.5}^{+4.8}$ & 19.2 & $0.19_{-0.18}^{+0.19}$ & $2.9_{-0.1}^{+1.2}$ & $6.2_{-0.9}^{+0.2} \times 10^{-4}$ & $1.0_{-1.0}^{+3.5}$ & $> 60.6$ & 1.03/535
\\
2017 Nov 25-27 & $7.1_{-1.1}^{+1.2}$ & $0.35_{-0.02}^{+0.03}$ & $2.8_{-0.6}^{+0.9}$ & 18.3 &   $-0.27_{-0.18}^{+0.04}$ & $2.9_{-0.2}^{+0.3}$ & $2.4_{-0.6}^{+0.1} \times 10^{-4}$ & $1.1_{-0.1}^{+1.7}$ & $> 21.6$ & 0.99/1078
\\
2019 Nov 18-22 & $6.7_{-3.1}^{+5.2}$ & $0.33_{-0.05}^{+0.15}$ & $0.6_{-0.4}^{+1.2}$ & 16.9 &  $-0.15_{-0.78}^{+0.43}$ & $3.0_{-1.3}^{+1.1}$ & $5.4_{-2.6}^{+0.6} \times 10^{-5}$ & $1.1_{-1.1}^{+0.0}$ & $> 32.4$  & 0.88/304
\\
\hline
\hline
\end{tabular}
\begin{flushleft}
{\footnotesize
Note: $^a$The percentage of the flux was measured from the best-fit within in 0.3--10\,keV.\\
}
\end{flushleft}
\end{table*}
\begin{figure*}[tp]
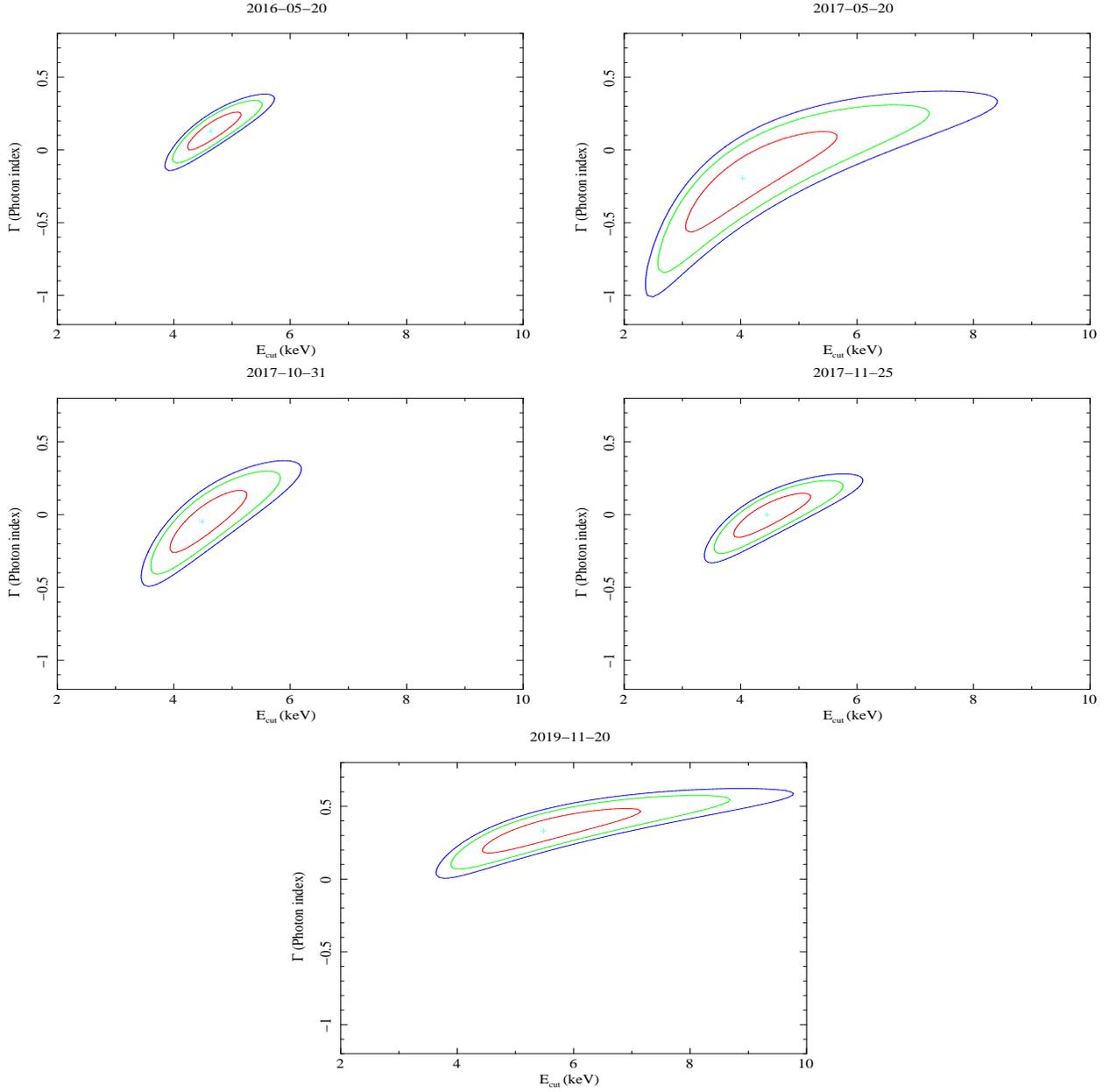

\centering
\includegraphics[width=5.5cm,height=8.5cm,angle=-90]{2016-05-20.eps}
\includegraphics[width=5.5cm,height=8.5cm,angle=-90]{2017-05-20.eps}
\includegraphics[width=5.5cm,height=8.5cm,angle=-90]{2017-10-31.eps}
\includegraphics[width=5.5cm,height=8.5cm,angle=-90]{2017-11-25.eps}
\includegraphics[width=5.5cm,height=8.5cm,angle=-90]{2019-11-20.eps}
\caption{{\footnotesize The contour plot of $\Gamma$ vs. $E{\rm cut}$ for the pulsed spectra of \psr\ at different epochs. We present the 1$\sigma$, 90\%  and 2$\sigma$ confidence contours for two parameters of interest in red, green and blue colors, respectively.}} 
\label{pspectrum}
\end{figure*}

\subsection{{\sl Spectral analysis}}
\label{ssec:Sanalysis}

To estimate the flux variation of our target at different epochs, we performed spectral analyses on individual data sets.
Since our target is not in the Milky Way, we therefore included a fixed Galactic absorption column ($3.5\times10^{20}$\,cm$^{-2}$; \citealt{HI4PI2016}) and a free intrinsic absorption column ($N_{\rm H,int}$) for an object in NGC~7793.
We applied the Tuebingen-Boulder ISM absorption model with an updated method to calculate the ISM cross section \citep{WAM2000} in the spectral fit. 
In the preliminary spectral fits, we found that a single component model is sufficient for an acceptable fit when our target stays in the low flux state.
We therefore applied a power-law with a high-energy exponential roll-off (i.e., CUTOFFPL model) to account for the presence of a curvature $\gtrsim 2$-4\,keV in the low flux states.
The single component model for the spectral fit is very poor in the high flux states, and an additional thermal component (i.e., DISKBB model; \citealt{Mitsuda84}) is included in our fit to provide a major contribution in the soft X-ray band ($\lesssim 2$\,keV).  
Following the absorption column measured in \citet{Walton2018}, we fixed the combined $N_{\rm H}$ (including both $N_{\rm H,Gal}$ and $N_{\rm H,int}$) as $10^{21}$\,cm$^{-2}$ to investigate the related spectral parameters at different epochs to derive the source flux within 3--10\,keV in Table~\ref{tsolution}. 

Some X-ray data have overlapping observation periods, and these observations provide us an opportunity to investigate the broad-band (0.5--20\,keV) spectral behavior.
We notice that the \emph{NuSTAR} data on 2019-11-18 has a duration of more than 1 day, and it is close to the starting time of \emph{XMM-Newton} observation on 2019-11-22.
We therefore also included these data sets in the broad-band spectral investigation.
To account for the cross-calibration mismatch between different X-ray detectors, we also introduced multiplicative scaling factors in our joint spectral fit.
In the preliminary fit to the broad-band spectra, we applied a composite spectral model of DISKBB+CUTOFFPL and a clear hard X-ray excess can be seen in the residuals at high-energies (i.e., $\gtrsim 15$\,keV).
The detection of such a hard X-ray excess in spectral fit was also found in \citet{Walton2018}, hence we follow the same idea to include one more component `SIMPL' convolved with the CUTOFFPL model to account for a fraction of the Comptonization photons scattered into the high-energy band \citep{Steiner2009}.
Table~\ref{broadband-spectrum} shows the evolution of spectral parameters and the 90\% uncertainties from the best fits in 5 different epochs since 2016.

In order to investigate the pulsed emission in the broad-band spectra, we further investigate the phase-resolved spectroscopy contributed by the pulsed component.
Following the timing solution of the pulsation detected at different epochs shown in Table~\ref{tsolution} and the FWHM determined for each pulse profile shown in Table~\ref{FWHM}, we can extract the `on-pulse' photons accordingly.
The `off-pulse' photons in the DC level can be extracted from the phase interval equivalent to the FWHM of pulsation, and we generated pulsed spectra by subtracting the `off-pulse' spectra from the `on-pulse' spectra.
Each channel except the boundary in the spectra were constrained to have at least 25 photons. 
We used an exponential cut-off power-law to describe the main contribution of the pulsed emission from the pulsar, and a high-energy excess can also be seen during the high-flux state.
However, we note that the cut-off power-law model has already provided acceptable fits to these pulsed spectra.
To evaluate the change of the spectral parameters, we fixed the intrinsic absorption column ($N_{\rm H,int}$) as 0 since the values obtained from all the best fits are small. 
Fig.~\ref{pspectrum} demonstrates the evolution of the photon index and cut-off energy in different stages.
We find that the pulsed spectrum is the hardest in mid-2017 with a relatively smaller cut-off energy, and it is somewhat different compared to the observations taken near the end of 2019.

\begin{figure}[tp]
\centering
\includegraphics[width=8.2cm,height=5.6cm]{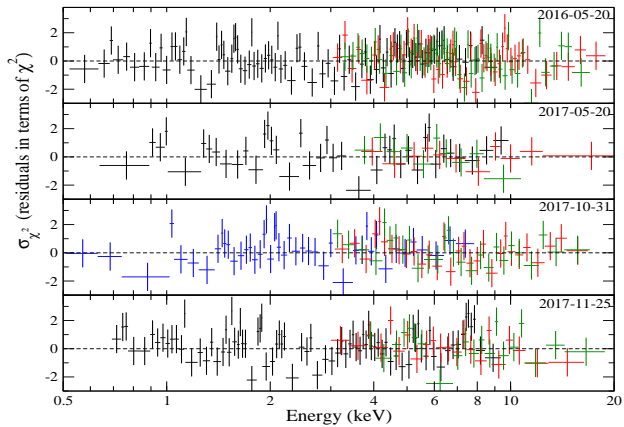}
\caption{{\footnotesize Residuals obtained from the pulsed spectral fit to the cut-off power-law model. We performed a joint fit to the spectra with a cut-off power-law at 4 different epochs labelled in each panel, and an absorption-like feature can be seen at 1.1--1.3\,keV from all the residual maps. Here we considered the spectra collected from \emph{XMM-Newton}/pn (black), \emph{NICER}/XTI (blue), \emph{NuSTAR}/FPMA (red) and \emph{NuSTAR}/FPMB (green).}} 
\label{residual}
\end{figure}

Except for the possible spectral change of the pulsed spectra for \psr, we also found an indication of a narrow absorption line feature at $\sim$1.3\,keV as shown in Fig.~\ref{residual}.  
This feature is not obvious in the phase-averaged spectra, but it is clearer in the pulsed spectra, especially in the high-flux stage.
We show this absorption line feature in Fig.~\ref{line} by performing a joint fit to four pulsed spectra collected at 2016 and 2017 using the CUTOFFPL model convolved with a Gaussian absorption line (gabs model in Xspec), and the line energy is at $1.26_{-0.04}^{+0.05}$\,keV with the uncertainties of 90\% confidence level.    
We also performed a likelihood ratio test with 10000 iterations to investigate the significance of this additional absorption feature, and only 0.3\% simulations can exceed the observed test statistic value, meaning that the significance is $\sim$3$\sigma$.   
The absorption-like feature around 1\,keV seen in several ULXs was interpreted as collisionally ionized gas or outflowing photoionized gas \citep{Middleton2014,Pinto2017}; however, such a feature is not significant in our detection when the unpulsed component dominates the spectra.
Therefore, the absorption line in our detection might be associated with the cyclotron resonant scattering feature (CRSF) in magnetized neutron stars \citep{CW2012,Staubert2019}.  

\section{{\sl Discussion}}
\label{sec:Discussion}

We used the archival \emph{Swift}, \emph{XMM-Newton}, \emph{NICER}, and \emph{NuSTAR} data to investigate the X-ray timing and spectral behaviors of \psr.
We revisited the pulsed detection and spectral flux of each observation and the evolved track is similar to that published in \citet{Furst2021}.
The flux and the PFs obtained from the individual data set are not totally in agreement with previous report because of the different selection of the background, different model for the spectral fits or different number of bins applied to fold the light curve.

\subsection{{\sl Pulse structure change}}
\label{ssec:Pchange}

Based on the support of finite GMM and two-sample Kuiper test, we further confirm the variability of the pulsation with a single-peaked structure. 
From mid-2016 to mid-2017, we find the narrowing of the pulsed emission, and then the major peak gradually increased the width after 2017 mid-May.
The pulsation remained broad until the end of 2019; however, the major peak was especially wide from the end of 2018 to mid-2019 and the pulsed structure at this time interval is different from that at the end of 2017 or of 2019 since two Gaussian components are required to describe the pulsed component.  
While \psr\ continuously decreased its flux level to one over fifth and kept a high PF (i.e., $\gtrsim 30$\% for both \emph{XMM-Newton} and \emph{NuSTAR} observations) during this time interval, the broader pulsation obtained in the similar time range may potentially indicate the change of the emission geometry which was led by less obscuration from the weak accretion.
In comparison to the similar pulsed structure detected with a low pulsed fraction/detection (i.e., PF\,$\lesssim 20$\%; $Z_1^2 \lesssim 100$) from 2017 mid-May to early July, the profile of the pulsation can be different with a strong pulsed fraction/detection depending on different accretion stages.

\begin{figure}[tp]
\centering
\includegraphics[width=5.6cm,height=8.8cm,angle=-90]{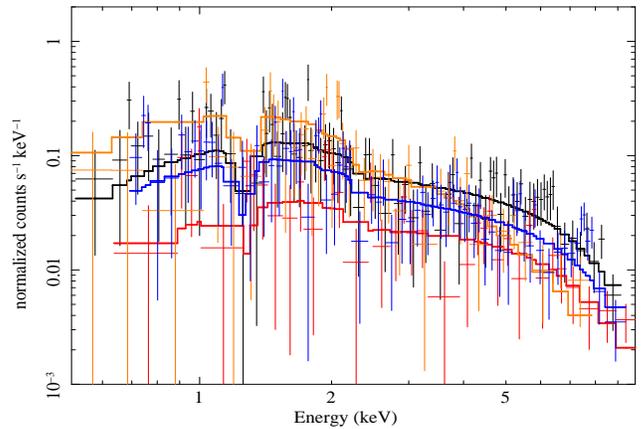}
\caption{{\footnotesize An absorbed line at $\sim$1.3\,keV obtained from the pulsed spectra of \emph{XMM-Newton}/pn, \emph{NICER}/XTI and \emph{NuSTAR}/FPM. We performed a joint fit to determine the absorption line feature from the pulsed spectra of 2016-05-20 (black), 2017-05-20 (red), 2017-10-31(orange) and 2017-11-25 (blue).}} 
\label{line}
\end{figure}

\subsection{{\sl Relation between the variation of the pulsed fraction and the orbital phase}}
\label{ssec:Variation_PF_OP}

\begin{figure}[tp]
\centering
\includegraphics[width=8.5cm,height=6.5cm]{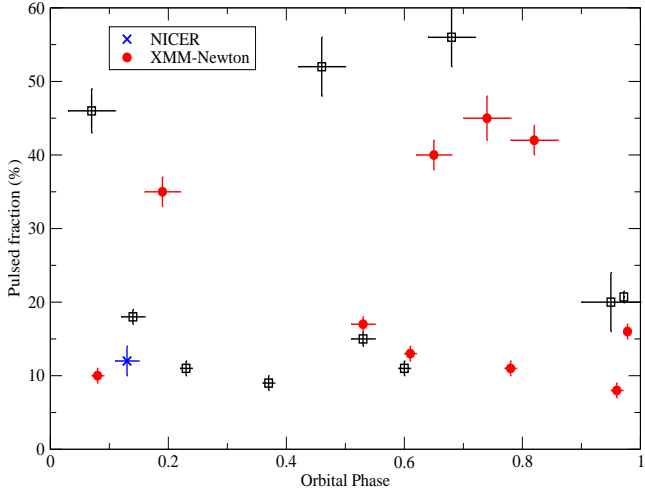}
\caption{{\footnotesize Correlation between the background-subtracted pulsed fraction in the 3--10\,keV band and the orbital phase of \psr. The orbital phase was inferred from the elliptical orbit determined in Table~\ref{orbitpara}. \emph{NuSTAR} data are shown in black squares, \emph{NICER} observation is labelled by blue cross, while \emph{XMM-Newton} observations are denoted by red circles.}} 
\label{PF_cirphase}
\end{figure}

The variation of the PF can be caused by the obscuration of the accretion flow or the disk wind from the neutron star at a specific position in the orbit.
We therefore also compute the related orbital phase of each data set based on timing parameters of an elliptical orbit determined in Table~\ref{orbitpara}.
Because the pulsed emission dominates in the hard X-ray band, we only compare the orbital phase with the background-subtracted PF in the 3--10\, keV band, which is the overlapped energy range for \emph{XMM-Newton}, \emph{NICER} and \emph{NuSTAR} observations.
To avoid serious contamination of the binning effect, here we counted the background-subtracted PF based on the Fourier decomposition \citep{DKG2009,HNH2019}.
No clear correlation can be found between the orbital phase and the PF as shown in Fig.~\ref{PF_cirphase}, but we can find that all the detected fractions can be divided into two groups.
For those with PFs above 25\% were all detected between 2018 mid-November and 2020 mid-January, when the source underwent a drop in the X-ray flux before entering an apparent off-state in 2020.
If our determination of the orbital period is reliable, we can at least confirm that the obscuration of the pulsed emission or the change of the emission geometry does not depend on the position of the neutron star in the orbit.
However, we must note that the observed pulsation of \psr\ decreased significantly (or even disappeared) in the off-state after early 2020. It can be originated from a different mechanism in comparison to the low PF detected in the high flux state.

\subsection{{\sl Marginal pulsed detection in mid-August of 2020}}
\label{ssec:PDon2020-08}

A very marginal pulsation can be detected on 2020-08-22 via the Rayleigh test.
We note that timing parameters of this signal are close to those inferred from the timing solution provided in \citet{Furst2018}, but the fit is poor with the updated timing solution determined in \citet{Furst2021} as well as in our work.
Therefore, more observations are required to confirm the validity of this candidate signal.
We also notice that the pulsed structure inferred from this marginal signal is similar to other folded light curves determined at weak pulsation in 2017.
The reason for those less sinusoidal profiles to show high similarity can be attributed to the major contribution from the background (i.e., DC level), and it means that a relatively high PF of $31\pm 9$\% measured for this marginal signal is not reliable due to the binning effect.
The PF of this candidate signal can be significantly decreased (i.e., $13\pm 3$\%) if we assessed it according to the Fourier decomposition \citep{DKG2009,HNH2019}.

\subsection{{\sl Evolution of the PF and the spectral behavior}}
\label{ssec:Evo_spec}

\citet{Furst2021} have investigated the correlation between the PF and the source flux as well as the hardness ratio.
It is suggested that \psr\ has a relatively larger PF when it stays in the low flux states. 
Similar findings are also found in our broad-band spectral analysis if we assume the pulsed emission of the accretion column can be described by an exponential cut-off power-law \citep{Walton2018}.
In our broad-band spectral fits shown in Table~\ref{broadband-spectrum}, we found a relatively large fraction of flux contribution in the 0.3--10\,keV band from the CUTOFFPL component although the source became fainter in the end of 2019.
According to the phase-resolved spectral fits (see Fig.~\ref{pspectrum}), we obtained a similar pulsed flux of $\sim 2\times 10^{-12}$\,ergs\,s$^{-1}$\,cm$^{-2}$ in the 0.3--20\,keV energy band in  mid-2017 and in the end of 2019.
However, \psr\ is much brighter in mid-2017, and therefore the PF is low in the corresponding time interval.

On the contrary, no significant correlation can be found between the PF and the spectral hardness \citep{Furst2021}. 
In our spectral analysis, the DISKBB component originated from the thermal accretion disk dominates the soft X-ray band, and the pulsed emission of the accretion column dominates the hard X-ray band.
Therefore, we would expect the spectral softening related to the decrease of the PF when the hard accretion column showed a fewer contribution in the spectrum.
We indeed observe fewer hard X-ray pulsed photons and less hard X-ray excess in the spectra leading to a relatively low pulsed fraction/detection in the pulsed spectral fit of mid-2017; however, the large uncertainty of the photon index prohibits us to confirm the correlation.
As shown in Fig.~\ref{pspectrum}, it is not easy to distinguish the spectral hardening from mid-2016 to the end of 2017 because of the large uncertainties.
We find that the pulsed spectrum is relatively soft in the end of 2019, but the higher PF in the same period is more closely correlated with the low source flux. 
Here the increase of the PF can be explained as the contamination by the scattering of beaming effect of X-ray photons escaped from a small evacuated cone \citep{King2009, Middleton2015}, which constrains the X-rays emitted within a limited open angle \citep{Furst2021}.

According to the spectral parameters determined from broad-band spectral fits, we find that \psr\ has a very hard spectrum, which can be classified into a `hard ultra-luminous state' with  $kT_{\rm in; DBB} < 0.5$\,keV and $\Gamma_{\rm CPL} < 2$ \citep{Sutton2013}. 
It implies that the beamed hard X-rays from the source were observed with a low inclination angle if we model the observed X-ray properties with a funnel shaped wind. 
The cut-off energy ($E_{\rm cut;CPL}$) determined from the pulsed spectra ($\gtrsim 4$\,keV) is larger than that obtained from the broad-band spectral fit because we included an additional component (i.e., SIMPL) to absorb the contribution from hard X-rays.
In contrast to the spectral parameters determined at the end of 2019, the thermal disk seems to have a higher temperature and larger size before 2018 (see Table~\ref{broadband-spectrum}).
Higher contributions from the thermal component also explain the relatively low PF detected before 2018.
\psr\ has an indication to recover its flux to a weak but stable level in mid-2021, and therefore future observations to survey the disappearance/weakening of pulsed emission can also be interesting for investigating the physics of a ULP.

\subsection{{\sl Implication of the detected CRSF}}
\label{ssec:Implication_CRSF}

According to the evolution of the PF observed for \psr, the change of the emission geometry can be explained by the scattering effects from an evacuated cone \citep{Koliopanos2017,Furst2018}, but such a geometry will not occur on a highly magnetized neutron star since the strong magnetic pressure will terminate the accretion flow to form a funnel-like structure close to the central accretor.
Therefore, \citet{Walton2018} constrained a limiting magnetic field strength of $B \lesssim 6\times 10^{12}$\,G  for this source with super-Eddington accretion in the thick inner disk.
\citet{Furst2016} used a standard accretion disk \citep{GL79,DPS2015} with a measured period of 418\,ms and a spin-up rate of $\sim 3.5\times 10^{-11}$\,s\,s$^{-1}$ to estimate the strength of the magnetic field as $\sim 1.5\times 10^{12}$\,G.  
Though we obtained an updated timing solution of the spin period and its derivative in our studies, the derived magnetic field strength is essentially the same since all the parameters are in the same order.
However, our detection for the CRSF provides a direct way to measure the magnetic field strength close to the neutron star surface \citep{Schonherr2007}.

CRSFs, which reveal as absorption-like lines at a fundamental Landau energy and its integer multiple \citep{Araya97}, was produced by resonant scattering of photons off electrons moving perpendicular to the direction of the magnetic field.
Since the energies quantized on Landau levels directly depend on the local magnetic field, the detected line energy of the CRSF ($E_{cyc}$) offers a direct way to trace the magnetic field strength assuming the scattering is given by the electrons.
 \begin{eqnarray}\label{eqno1}
E_{cyc} =\frac{n}{1+z}11.6[{\rm keV}]\times B_{12},
\end{eqnarray}
where $n$ is the Landau levels, $z$ is the gravitational redshift of a neutron star, and $B_{12}$ is the strength of the magnetic field scaled in $10^{12}$\,Gauss \citep{Staubert2019}.      

Here we consider the case of fundamental line for a scattering from the ground level to the first excited Landau level to infer the largest magnetic field, and we determine the surface gravitational redshift as $z=0.3$ for a canonical neutron star with a typical size \citep{Liang86}.  
With Eq. 1, we can obtain a surface magnetic field strength of $\sim 1.5\times 10^{11}$\,G if the absorption line detected at $\sim 1.3$\,keV in Fig.~\ref{line} can indeed denote a CRSF.
We notice that the centroid line energy in our detection is an order smaller than other usual cases (e.g.,  \citealt{Furst2014}, \citealt{Staubert2014}, and \citealt{Furst2015}), and it leads to the classification of the magnetic field strength to our target as a relatively low-$B$ neutron star \citep{DPS2015}.  
In order to have a super-Eddington accretion from a low-$B$ neutron star, a small beaming factor ($\lesssim 0.2$) is also required in this system \citep{King2008}. 
We note that such a line feature detected below 10\,keV can also correspond to a proton cyclotron line detected in a magnetar \citep{Ibrahim2002,ISP2003}.
The inferred magnetic field from such a proton cyclotron line is in an order of $> 10^{14}$\,G, which is significantly deviated from the magnetic field strength estimated from the timing parameters.
We note that the mismatch of the magnetic field strength inferred from the timing parameters and proton cyclotron lines was discussed in \citet{Tiengo2013}. Nevertheless, such a strong magnetic field may be controversial.
It is more likely that a relatively low magnetic field strength ($\sim 10^{11}$\,G) induces a CRSF if our detection is real.   

A better understanding of the complex shape for this absorption feature can reveal the physics of the emitting process and the scattering geometry in accreting pulsars \citep{Schwarm2017I, Schwarm2017II}. Unfortunately, the signal-to-noise of this feature is not sufficient for us to precisely constrain other parameters (e.g., line depth).
Except for the source went into the low (pulsed) flux state, we can always find evidence of this absorption feature. 
We only yield a $\sim$3$\sigma$ significance through the likelihood ratio test from 4 spectra with obvious line detection.
Because we only detect the marginal CRSF from the pulsed spectra, the strength of CRSF seems to correlate with the pulsed phase and such an interesting behavior was also discovered for other sources (e.g., KS 1947+300; \citealt{Furst2014}).
In addition, we also find an indication of a small variation of the line energy at different time epochs, and such a variation can be due to the X-ray luminosity, spin pulsation or the superorbital modulation of the system (e.g., Hercules X-1; \citealt{Vasco2013,Staubert2014}).
Unfortunately, the current data sets do not allow us to carry out any further detailed investigations on this feature, and future X-ray observations with long exposures and better energy resolution are required to help us clarify the line detection.

\section{{\sl Summary}}
\label{sec:Summary}

We performed timing and spectral analyses with the X-ray archival data taken since mid-2016 to study the ULP, \psr\, evolving from the luminous to the faint state with an order of magnitude in flux change. In the following we briefly summarize the obtained results.  
 
\begin{enumerate}[label=\textbf{\arabic*.}]
\setlength\itemsep{0.1em}
\item
We statistically compared the pulsed structure, and there were at least three obvious changes in the distribution of the profile during the time between mid-2016 and mid-2017, between mid-2017 and the end of 2017, and between the end of 2017 and the end of 2018. According to the track of the flux evolution, the onset of the changing pulsed structure seems to correlate with the flux evolution. 
\item
An investigation for the correlation of the PF and the orbital period ($\sim$65\,d) suggests that there is no clear relation between the PF change and the orbital phase. Generally speaking, the measured PF of \psr\ in 3--10\,keV can be divided into two groups. 
\item
The pulsation became very weak after mid-2020, but a hint of the detection is evident.
\item
\psr\ always stayed in the hard ultra-luminous state during our investigations. Nevertheless, the pulsed emission seems to become relatively softer to lead a higher pulsed contribution in a few keV when the source went into a faint stage since the end of 2019. 
\item
An absorption feature at $\sim$1.3\,keV potentially corresponding to CRSF is marginally detected in the pulsed spectra. 
\end{enumerate} 

We note that X-ray monitoring of \psr\ is on-going, and our discoveries can be verified in the future. More interesting features of this ULP are also expected to investigate the accretion mechanism of this source. 

\acknowledgments
This work made use of archival data provided by the database of NASA's High Energy Astrophysics Science Archive Research Center (HEASARC).
This work is supported by the National Research Foundation of Korea (NRFK) through grant 2016R1A5A1013277 and by the Ministry of Science and Technology (MoST) of Taiwan through grant MOST 110-2811-M-006-515 and MOST 110-2112-M-006-006-MY3.
C.-P.~H. acknowledges support from the MoST of Taiwan through grant MOST 109-2112-M-018-009-MY3.
J.~T. is supported by the National Key Research and Development Program of China (grant No. 2020YFC2201400) and the National Natural Science Foundation of China (grant No. U1838102).
K.-L.~L. is supported by the MoST of Taiwan through grant 110-2636-M-006-013, and he is also a Yushan (Young) Scholar of the Ministry of Education of Taiwan.
C.~Y.~H. is supported by the National Research Foundation of Korea through grant 2016R1A5A1013277 and 2019R1F1A1062071.
A.~K.~H.~K. is supported by the Ministry of Science and Technology (MoST) of Taiwan through grants 108-2628-M-007-005-RSP and 109-2628-M-007-005-RSP.

{\it Facilities:} \facility{{\it Swift}(XRT), {\it XMM-Newton}(EPIC), {\it NICER}(XTI), and {\it NuSTAR}(FPM), HEASARC}. 

\software{\newline {\tt Science Analysis System} \\ (\url{https://www.cosmos.esa.int/web/xmm-newton/how-to-use-sas}; \citealt{SAS2004})
\newline {\tt HEASoft} \\ (\url{https://heasarc.gsfc.nasa.gov/docs/software/lheasoft/\\developers\_guide/}; \citealt{HEAsoft2014})
\newline {\tt Xspec} \\ (\url{https://heasarc.gsfc.nasa.gov/xanadu/xspec/}; \citealt{Xspec96})
\newline {\tt Mclust} \\ (\url{https://cran.r-project.org/web/packages/mclust/vignettes/\\mclust.html}; \citealt{Xspec96})
}

\appendix
\section{A. Finite Gaussian mixture modelling}\label{app:GMM}

To model the variety of random phenomenon for clustering, classification and density estimation, more and more astronomers consider the related investigations with finite mixture models \citep{FMM}, which are computationally convenient to model complex distributions of data.
\texttt{CRAN Mclust} package (version  5.4.6; \citealt{Scrucca2016}) is a powerful R package, which includes hierarchical clustering, expectation-maximization algorithm for mixture estimation and different tools for model selection, to model the data with a Gaussian finite mixture.
We note that the most updated version considers the variety of covariance structures through eigenvalue decomposition and counts for different number of mixture components to satisfy our motivation to differentiate the pulsed structural change by the clustering of photon arrival phases.
We therefore perform a density estimation on the unbinned distribution of the rotational phases \citep{Lin2021} determined by the local timing ephemeris shown in Table~\ref{tsolution}. 

\begin{figure*}[b]
\centering
\includegraphics[width=4.4cm,height=5.8cm]{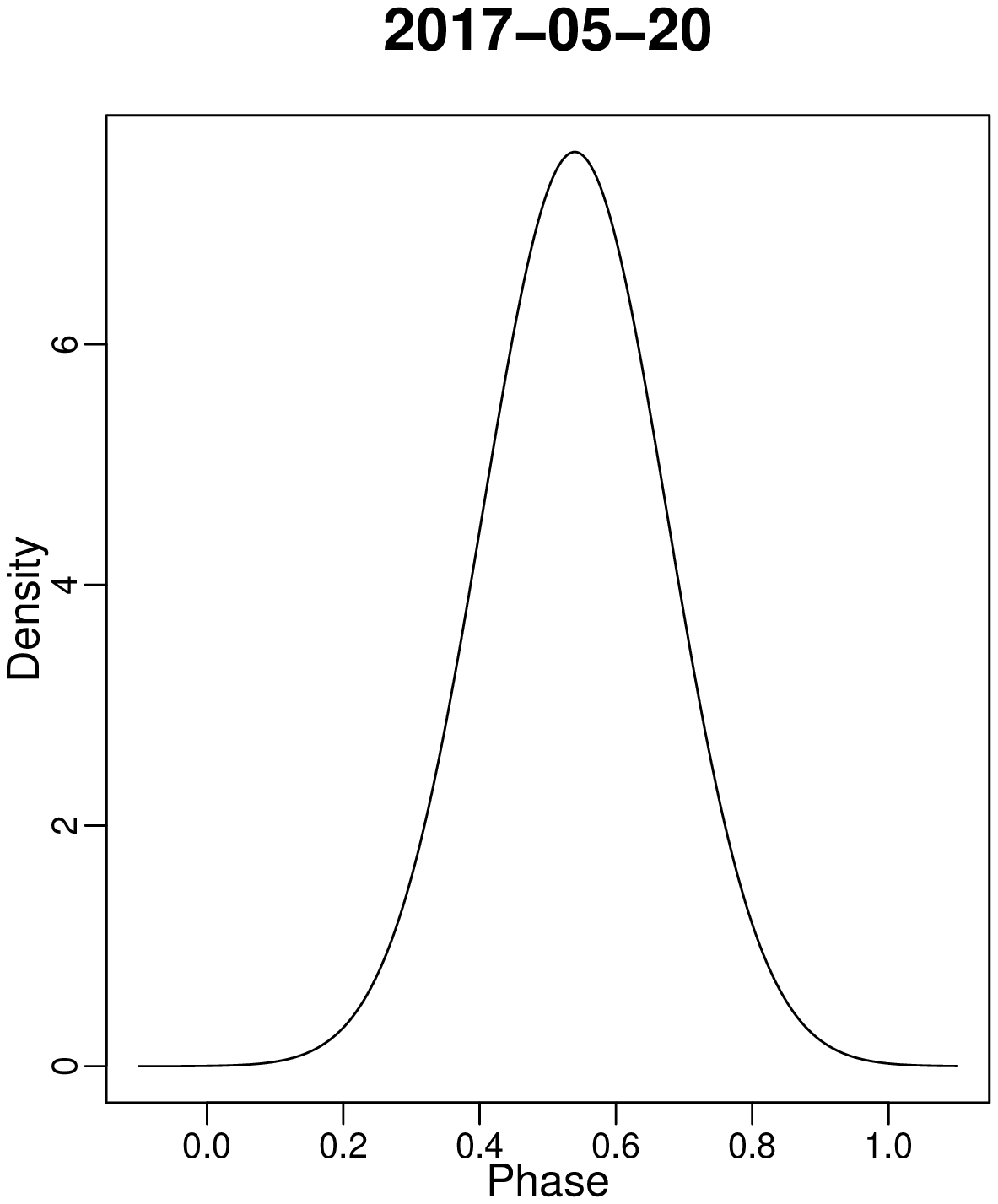}
\includegraphics[width=4.4cm,height=5.8cm]{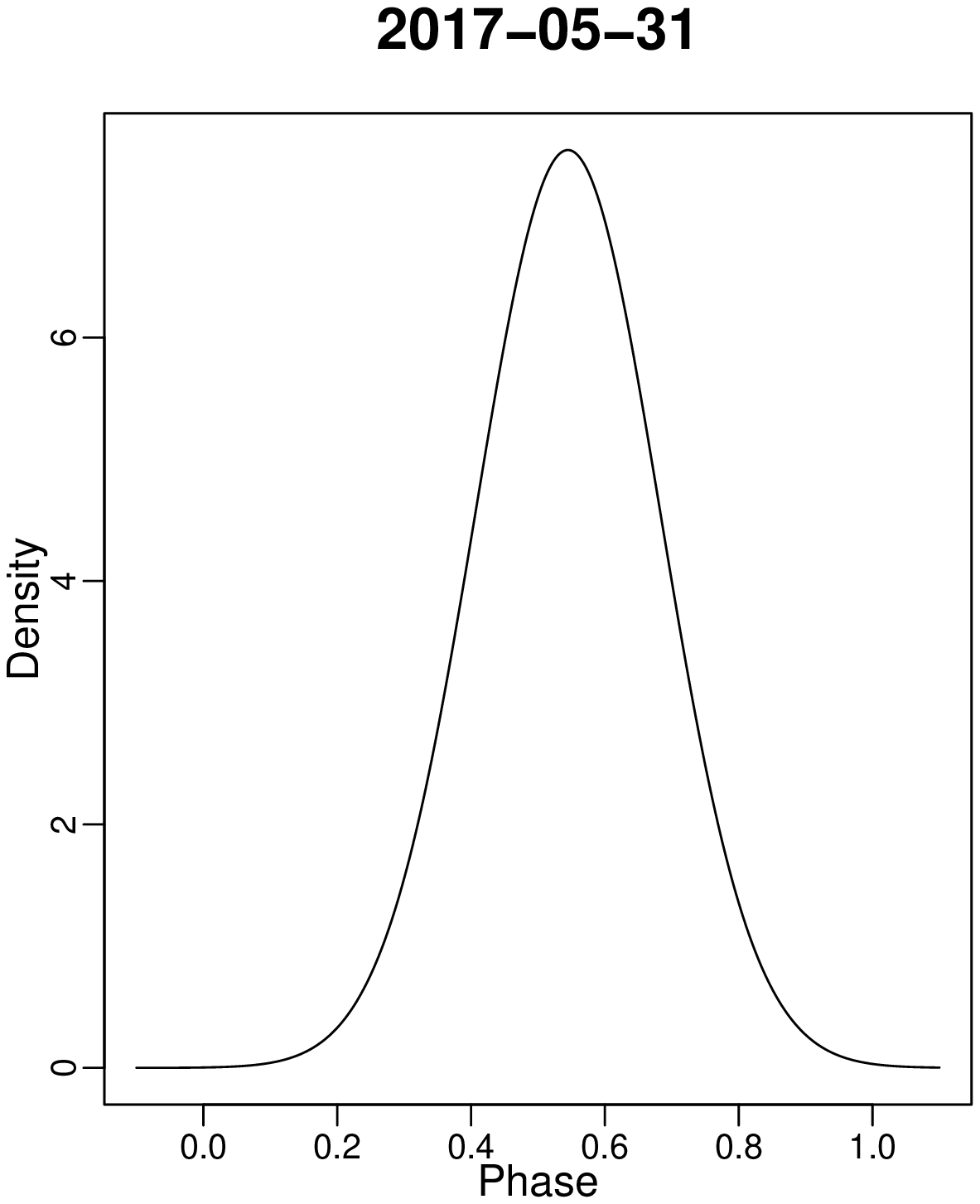}
\includegraphics[width=4.4cm,height=5.8cm]{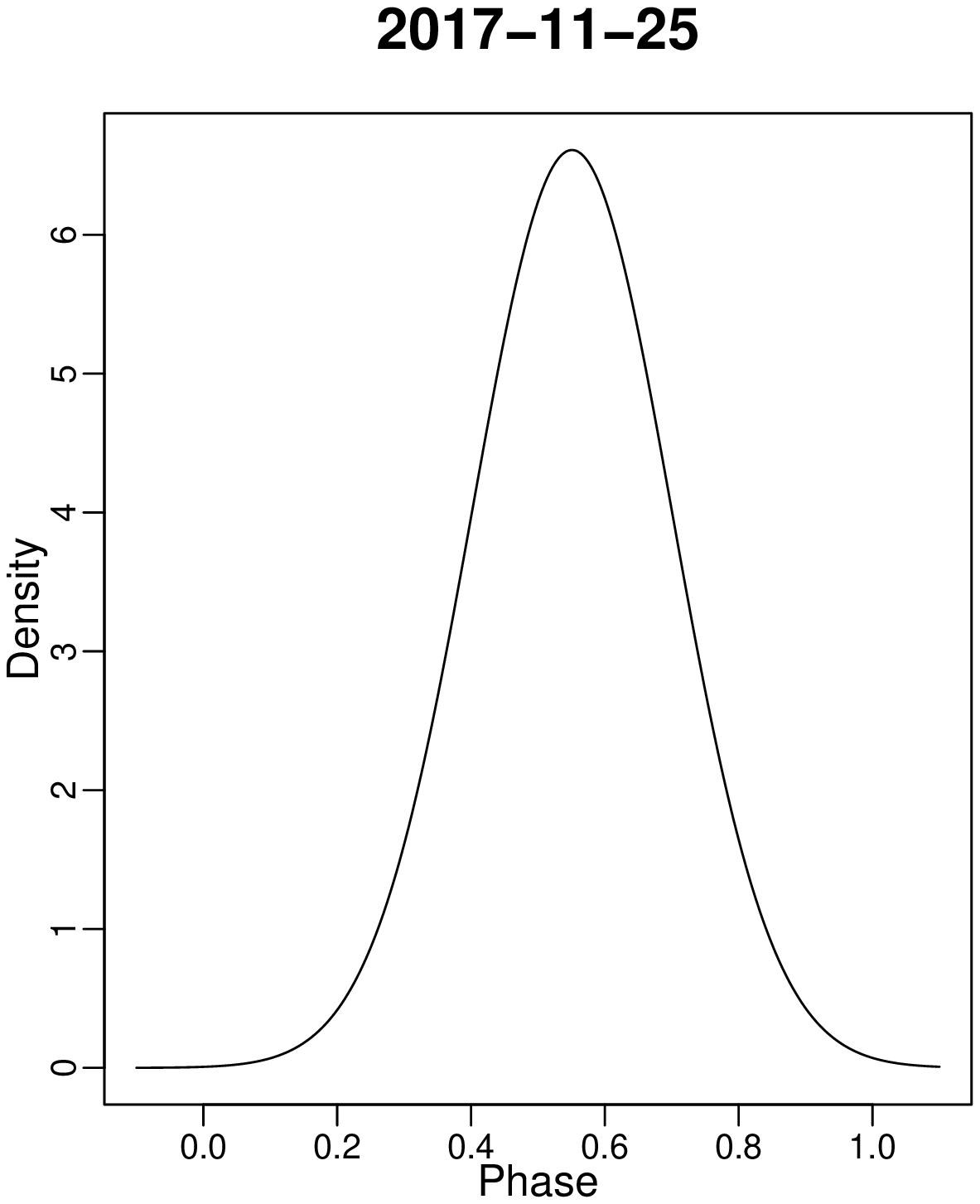}
\includegraphics[width=4.4cm,height=5.8cm]{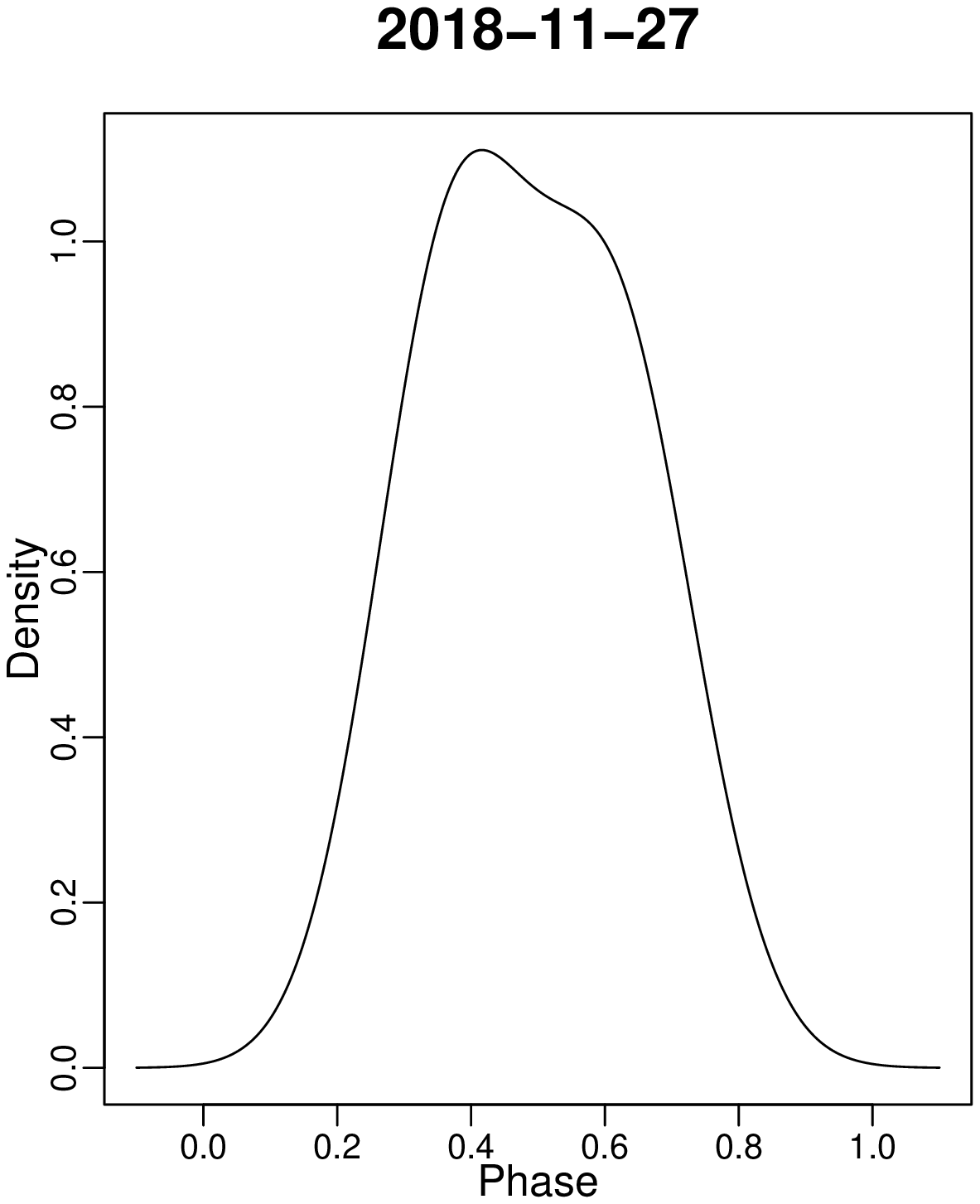}
\caption{{\footnotesize Density profiles estimated by fitting the GMM to the unbinned rotational phases and the BIC as a function of the number of components. Here we present an example of 4 profiles obtained from the pulsation detected for \emph{XMM-newton} observations in Table~\ref{tsolution} at 4 different epochs. The position of the strongest peak is shifted to the phase around 0.5, and the FWHM of each Gaussian component is presented in Table~\ref{FWHM}. }} 
\label{density}
\end{figure*}

The density of the arrival phase obtained at different time epochs was fitted using \texttt{CRAN Mclust} package, which comprises 1--8 Gaussian components and the Poisson noise to model the fluctuation of the unpulsed photons. 
The Bayesian information criterion (BIC; \citealt{Jackson2005}) was used as a function of number of components to describe the best-fit density profile, and we show some examples in Fig.~\ref{density}.
We note that BIC is one usual choice in the context of GMMs, and takes the form in the the log-likelihood:  $BIC_{M}=2\ln L\left(x|M,\hat{\theta}\right) - \nu\ln N$, where $L$ is the log-likelihood function at the maximum likelihood estimate $\hat{\theta}$ for the model $M$ in the presence of observed data $x$. 
$\nu$ and $N$ are the number of free parameters and sample size respectively. 

\begin{figure*}
\centering
\includegraphics[width=8.5cm,height=6.5cm]{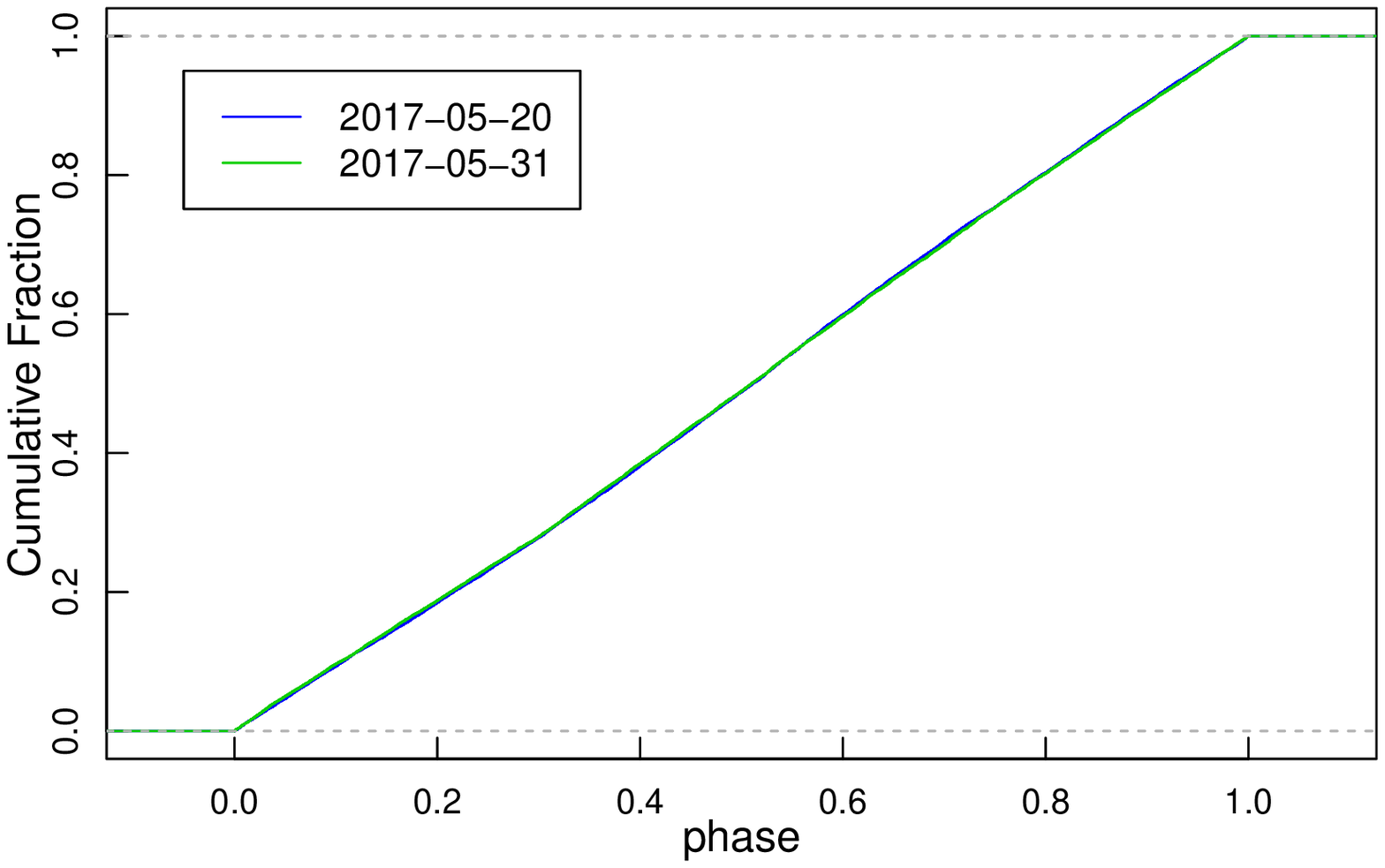}
\includegraphics[width=8.5cm,height=6.5cm]{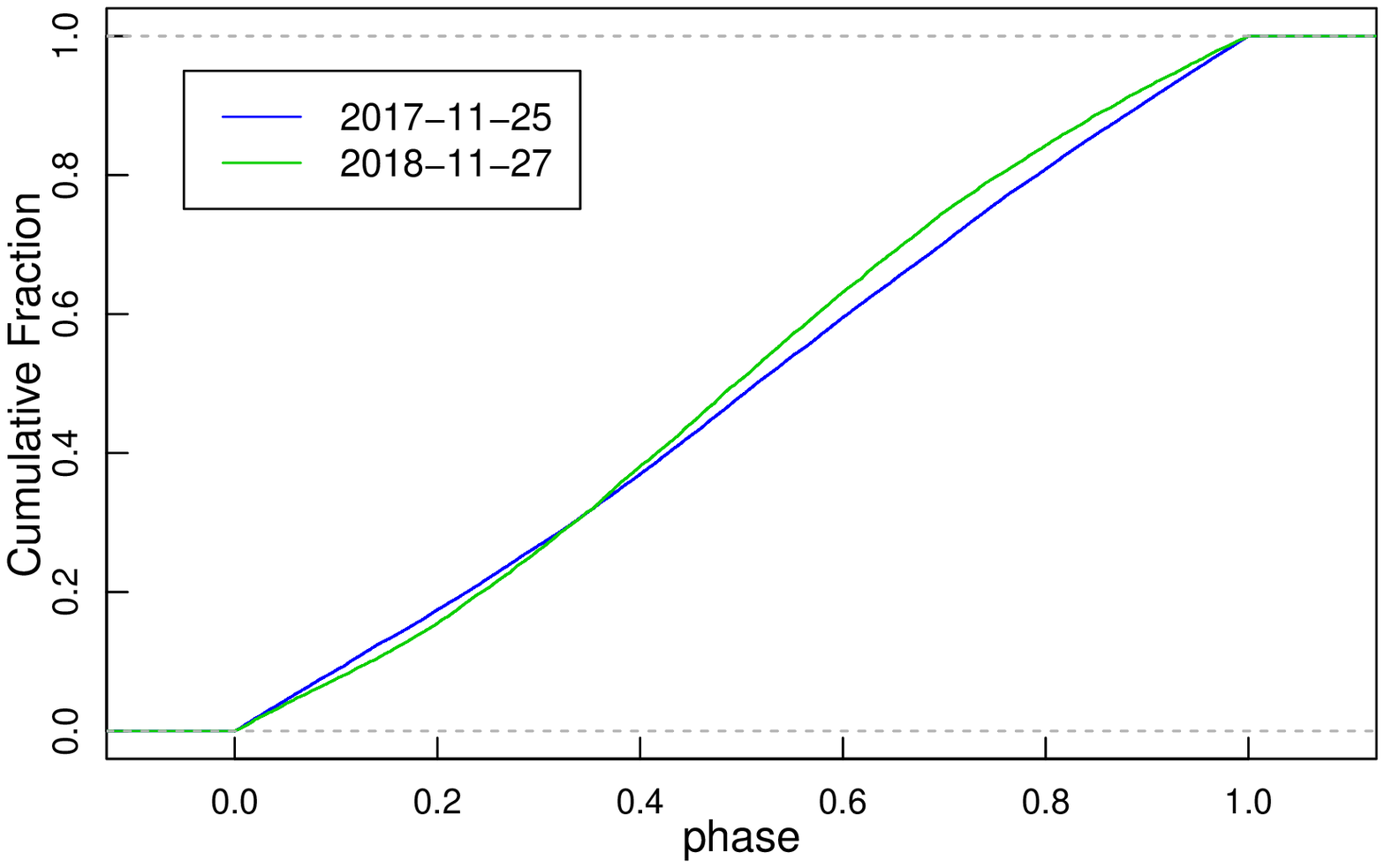}
\caption{{\footnotesize Cumulative distributions of the rotational phase for two unbinned pulse profiles determined at different epochs. Here we demonstrate two examples to compare the pulsed structure by two-sample Kuiper test. The associated null hypothesis probability obtained from the test can be referred to Table~\ref{KS-test}. The left panel indicates a high similarity of two pulse profiles, while the right panel shows a case resulted from two profiles of different structures.}} 
\label{fraction}
\end{figure*}

\section{B. Two-sample Kuiper test}\label{app:Kuiper-test}

Kuiper test is generally applied to confirm whether the small change of the pulsed structure is true or not (e.g., \citealt{Clark2017}).
To statistically investigate the variation of the pulsation, we computed the distribution of the arrival phase for each photon according to the spin frequency and its time derivative at different epochs reported in Table~\ref{tsolution}.
Then we shifted the arrival phases obtained at two different epochs with a resolution of 0.01 and searched for the minimum test statistic so as to avoid the false alarm from the misaligned main peak.
The Kuiper test statistic ($V_{n_1,n_2}$) can be calculated by $V_{n_1,n_2}=\max{\left\{F_{n_1}(\phi) - F_{n_2}(\phi) \right\}} + \max{\left\{F_{n_2}(\phi) - F_{n_1}(\phi) \right\}}$, where $F_{n_1}(\phi)$ and $F_{n_2}(\phi)$ denote two different cumulative distribution functions of phase with $n_1$ and $n2$ data points.
The null hypothesis is that the rotational phase distributions corresponding to the pulse profile resulted from two independent time segments are similar.
The ``Prob.'' in Table~\ref{KS-test} indicates the probability (or p-value; \citealt{JP96}) to obtain a two-sample Kuiper test statistic larger than or equal to the observed value under the null hypothesis, and Fig.~\ref{fraction} illustrates two examples of our data investigated by the two-sample Kuiper test.


\begin{thebibliography}{81}
\expandafter\ifx\csname natexlab\endcsname\relax\def\natexlab#1{#1}\fi

\bibitem[{{Araya}(1997)}]{Araya97}
{Araya}, R.~A. 1997, PhD thesis, THE JOHNS HOPKINS UNIVERSITY

\bibitem[{{Arnaud}(1996)}]{Xspec96}
{Arnaud}, K.~A. 1996, in Astronomical Society of the Pacific Conference Series,
  Vol. 101, Astronomical Data Analysis Software and Systems V, ed. G.~H.
  {Jacoby} \& J.~{Barnes}, 17

\bibitem[{{Arzoumanian} {et~al.}(2014){Arzoumanian}, {Gendreau}, {Baker},
  {Cazeau}, {Hestnes}, {Kellogg}, {Kenyon}, {Kozon}, {Liu}, {Manthripragada},
  {Markwardt}, {Mitchell}, {Mitchell}, {Monroe}, {Okajima}, {Pollard},
  {Powers}, {Savadkin}, {Winternitz}, {Chen}, {Wright}, {Foster}, {Prigozhin},
  {Remillard}, \& {Doty}}]{Arzoumanian2014}
{Arzoumanian}, Z., {et~al.} 2014, in Society of Photo-Optical Instrumentation
  Engineers (SPIE) Conference Series, Vol. 9144, Space Telescopes and
  Instrumentation 2014: Ultraviolet to Gamma Ray, ed. T.~{Takahashi}, J.-W.~A.
  {den Herder}, \& M.~{Bautz}, 914420

\bibitem[{{Atapin}(2018)}]{Atapin2018}
{Atapin}, K. 2018, in Accretion Processes in Cosmic Sources - II, 38

\bibitem[{{Bachetti} {et~al.}(2014){Bachetti}, {Harrison}, {Walton},
  {Grefenstette}, {Chakrabarty}, {F{\"u}rst}, {Barret}, {Beloborodov}, {Boggs},
  {Christensen}, {Craig}, {Fabian}, {Hailey}, {Hornschemeier}, {Kaspi},
  {Kulkarni}, {Maccarone}, {Miller}, {Rana}, {Stern}, {Tendulkar}, {Tomsick},
  {Webb}, \& {Zhang}}]{Bachetti2014}
{Bachetti}, M., {et~al.} 2014, \nat, 514, 202

\bibitem[{{Bachetti} {et~al.}(2021){Bachetti}, {Markwardt}, {Grefenstette},
  {Gotthelf}, {Kuiper}, {Barret}, {Cook}, {Davis}, {F{\"u}rst}, {Forster},
  {Harrison}, {Madsen}, {Miyasaka}, {Roberts}, {Tomsick}, \&
  {Walton}}]{Bachetti2021}
---. 2021, \apj, 908, 184

\bibitem[{{Brightman} {et~al.}(2020){Brightman}, {Earnshaw}, {F{\"u}rst},
  {Harrison}, {Heida}, {Israel}, {Pike}, {Stern}, \& {Walton}}]{Brightman2020}
{Brightman}, M., {et~al.} 2020, \apj, 895, 127

\bibitem[{{Caballero} \& {Wilms}(2012)}]{CW2012}
{Caballero}, I., \& {Wilms}, J. 2012, \memsai, 83, 230

\bibitem[{{Carpano} {et~al.}(2018){Carpano}, {Haberl}, {Maitra}, \&
  {Vasilopoulos}}]{Carpano2018}
{Carpano}, S., {Haberl}, F., {Maitra}, C., \& {Vasilopoulos}, G. 2018, \mnras,
  476, L45

\bibitem[{{Chandra} {et~al.}(2020){Chandra}, {Roy}, {Agrawal}, \&
  {Choudhury}}]{Chandra2020}
{Chandra}, A.~D., {Roy}, J., {Agrawal}, P.~C., \& {Choudhury}, M. 2020, \mnras,
  495, 2664

\bibitem[{{Clark} {et~al.}(2017){Clark}, {Wu}, {Pletsch}, {Guillemot}, {Allen},
  {Aulbert}, {Beer}, {Bock}, {Cu{\'e}llar}, {Eggenstein}, {Fehrmann}, {Kramer},
  {Machenschalk}, \& {Nieder}}]{Clark2017}
{Clark}, C.~J., {et~al.} 2017, \apj, 834, 106

\bibitem[{{Dall'Osso} {et~al.}(2015){Dall'Osso}, {Perna}, \&
  {Stella}}]{DPS2015}
{Dall'Osso}, S., {Perna}, R., \& {Stella}, L. 2015, \mnras, 449, 2144

\bibitem[{{de Jager}(1994)}]{Jager94}
{de Jager}, O.~C. 1994, \apj, 436, 239

\bibitem[{{Dib} {et~al.}(2009){Dib}, {Kaspi}, \& {Gavriil}}]{DKG2009}
{Dib}, R., {Kaspi}, V.~M., \& {Gavriil}, F.~P. 2009, \apj, 702, 614

\bibitem[{{Eksi} {et~al.}(2015){Eksi}, {Andac}, {Cikintoglu}, {Gencali},
  {Gungor}, \& {Oztekin}}]{Eksi2015}
{Eksi}, K.~Y., {Andac}, I.~C., {Cikintoglu}, S., {Gencali}, A.~A., {Gungor},
  C., \& {Oztekin}, F. 2015, \mnras, 448, L40

\bibitem[{{Evans} {et~al.}(2007){Evans}, {Beardmore}, {Page}, {Tyler},
  {Osborne}, {Goad}, {O'Brien}, {Vetere}, {Racusin}, {Morris}, {Burrows},
  {Capalbi}, {Perri}, {Gehrels}, \& {Romano}}]{Evans2007}
{Evans}, P.~A., {et~al.} 2007, \aap, 469, 379

\bibitem[{{Evans} {et~al.}(2009){Evans}, {Beardmore}, {Page}, {Osborne},
  {O'Brien}, {Willingale}, {Starling}, {Burrows}, {Godet}, {Vetere}, {Racusin},
  {Goad}, {Wiersema}, {Angelini}, {Capalbi}, {Chincarini}, {Gehrels}, {Kennea},
  {Margutti}, {Morris}, {Mountford}, {Pagani}, {Perri}, {Romano}, \&
  {Tanvir}}]{Evans2009}
---. 2009, \mnras, 397, 1177

\bibitem[{{Feng} \& {Soria}(2011)}]{FS2011}
{Feng}, H., \& {Soria}, R. 2011, Nature, 55, 166

\bibitem[{{F{\"u}rst} {et~al.}(2014){F{\"u}rst}, {Pottschmidt}, {Wilms},
  {Kennea}, {Bachetti}, {Bellm}, {Boggs}, {Chakrabarty}, {Christensen},
  {Craig}, {Hailey}, {Harrison}, {Stern}, {Tomsick}, {Walton}, \&
  {Zhang}}]{Furst2014}
{F{\"u}rst}, F., {et~al.} 2014, \apjl, 784, L40

\bibitem[{{F{\"u}rst} {et~al.}(2015){F{\"u}rst}, {Pottschmidt}, {Miyasaka},
  {Bhalerao}, {Bachetti}, {Boggs}, {Christensen}, {Craig}, {Grinberg},
  {Hailey}, {Harrison}, {Kennea}, {Rahoui}, {Stern}, {Tendulkar}, {Tomsick},
  {Walton}, {Wilms}, \& {Zhang}}]{Furst2015}
---. 2015, \apjl, 806, L24

\bibitem[{{F{\"u}rst} {et~al.}(2016){F{\"u}rst}, {Walton}, {Harrison}, {Stern},
  {Barret}, {Brightman}, {Fabian}, {Grefenstette}, {Madsen}, {Middleton},
  {Miller}, {Pottschmidt}, {Ptak}, {Rana}, \& {Webb}}]{Furst2016}
---. 2016, \apjl, 831, L14

\bibitem[{{F{\"u}rst} {et~al.}(2018){F{\"u}rst}, {Walton}, {Heida}, {Harrison},
  {Barret}, {Brightman}, {Fabian}, {Middleton}, {Pinto}, {Rana}, {Tramper},
  {Webb}, \& {Kretschmar}}]{Furst2018}
---. 2018, \aap, 616, A186

\bibitem[{{F{\"u}rst} {et~al.}(2021){F{\"u}rst}, {Walton}, {Heida}, {Bachetti},
  {{Pinto}, C. and {Middleton}, M.~J. and {Brightman}, M. and {Earnshaw}, H. P.
  and Barret}, {Fabian}, {Kretschmar}, {Pottschmidt}, {Ptak}, T., {Stern},
  {Webb}, \& {Wilms}}]{Furst2021}
---. 2021, Long-term pulse period evolution of the ultra-luminous X-ray pulsar
  NGC 7793 P13

\bibitem[{{Gabriel} {et~al.}(2004){Gabriel}, {Denby}, {Fyfe}, {Hoar}, {Ibarra},
  {Ojero}, {Osborne}, {Saxton}, {Lammers}, \& {Vacanti}}]{SAS2004}
{Gabriel}, C., {et~al.} 2004, in Astronomical Society of the Pacific Conference
  Series, Vol. 314, Astronomical Data Analysis Software and Systems (ADASS)
  XIII, ed. F.~{Ochsenbein}, M.~G. {Allen}, \& D.~{Egret}, 759

\bibitem[{{Ghosh} \& {Lamb}(1979)}]{GL79}
{Ghosh}, P., \& {Lamb}, F.~K. 1979, \apj, 232, 259

\bibitem[{{Gibson} {et~al.}(1982){Gibson}, {Harrison}, {Kirkman}, {Lotts},
  {Macrae}, {Orford}, {Turver}, \& {Walmsley}}]{Gibson82}
{Gibson}, A.~I., {Harrison}, A.~B., {Kirkman}, I.~W., {Lotts}, A.~P., {Macrae},
  J.~H., {Orford}, K.~J., {Turver}, K.~E., \& {Walmsley}, M. 1982, \nat, 296,
  833

\bibitem[{{Harrison} {et~al.}(2013){Harrison}, {Craig}, {Christensen},
  {Hailey}, {Zhang}, {Boggs}, {Stern}, {Cook}, {Forster}, {Giommi},
  {Grefenstette}, {Kim}, {Kitaguchi}, {Koglin}, {Madsen}, {Mao}, {Miyasaka},
  {Mori}, {Perri}, {Pivovaroff}, {Puccetti}, {Rana}, {Westergaard}, {Willis},
  {Zoglauer}, {An}, {Bachetti}, {Barri{\`e}re}, {Bellm}, {Bhalerao},
  {Brejnholt}, {Fuerst}, {Liebe}, {Markwardt}, {Nynka}, {Vogel}, {Walton},
  {Wik}, {Alexander}, {Cominsky}, {Hornschemeier}, {Hornstrup}, {Kaspi},
  {Madejski}, {Matt}, {Molendi}, {Smith}, {Tomsick}, {Ajello}, {Ballantyne},
  {Balokovi{\'c}}, {Barret}, {Bauer}, {Blandford}, {Brandt}, {Brenneman},
  {Chiang}, {Chakrabarty}, {Chenevez}, {Comastri}, {Dufour}, {Elvis}, {Fabian},
  {Farrah}, {Fryer}, {Gotthelf}, {Grindlay}, {Helfand}, {Krivonos}, {Meier},
  {Miller}, {Natalucci}, {Ogle}, {Ofek}, {Ptak}, {Reynolds}, {Rigby},
  {Tagliaferri}, {Thorsett}, {Treister}, \& {Urry}}]{Harrison2013}
{Harrison}, F.~A., {et~al.} 2013, \apj, 770, 103

\bibitem[{{HI4PI Collaboration} {et~al.}(2016){HI4PI Collaboration}, {Ben
  Bekhti}, {Fl{\"o}er}, {Keller}, {Kerp}, {Lenz}, {Winkel}, {Bailin},
  {Calabretta}, {Dedes}, {Ford}, {Gibson}, {Haud}, {Janowiecki}, {Kalberla},
  {Lockman}, {McClure-Griffiths}, {Murphy}, {Nakanishi}, {Pisano}, \&
  {Staveley-Smith}}]{HI4PI2016}
{HI4PI Collaboration} {et~al.} 2016, \aap, 594, A116

\bibitem[{{Horne} \& {Baliunas}(1986)}]{HB86}
{Horne}, J.~H., \& {Baliunas}, S.~L. 1986, ApJ, 302, 757

\bibitem[{{Hu} {et~al.}(2017){Hu}, {Li}, {Kong}, {Ng}, \& {Lin}}]{Hu2017}
{Hu}, C.-P., {Li}, K.~L., {Kong}, A. K.~H., {Ng}, C.~Y., \& {Lin}, L. C.-C.
  2017, \apjl, 835, L9

\bibitem[{{Hu} {et~al.}(2019){Hu}, {Ng}, \& {Ho}}]{HNH2019}
{Hu}, C.-P., {Ng}, C.~Y., \& {Ho}, W. C.~G. 2019, \mnras, 485, 4274

\bibitem[{{Hu} {et~al.}(2021){Hu}, {Ueda}, \& {Enoto}}]{HUE2021}
{Hu}, C.-P., {Ueda}, Y., \& {Enoto}, T. 2021, \apj, 909, 5

\bibitem[{{Ibrahim} {et~al.}(2002){Ibrahim}, {Safi-Harb}, {Swank}, {Parke},
  {Zane}, \& {Turolla}}]{Ibrahim2002}
{Ibrahim}, A.~I., {Safi-Harb}, S., {Swank}, J.~H., {Parke}, W., {Zane}, S., \&
  {Turolla}, R. 2002, \apjl, 574, L51

\bibitem[{{Ibrahim} {et~al.}(2003){Ibrahim}, {Swank}, \& {Parke}}]{ISP2003}
{Ibrahim}, A.~I., {Swank}, J.~H., \& {Parke}, W. 2003, \apjl, 584, L17

\bibitem[{{Israel} {et~al.}(2017{\natexlab{a}}){Israel}, {Belfiore}, {Stella},
  {Esposito}, {Casella}, {De Luca}, {Marelli}, {Papitto}, {Perri}, {Puccetti},
  {Castillo}, {Salvetti}, {Tiengo}, {Zampieri}, {D'Agostino}, {Greiner},
  {Haberl}, {Novara}, {Salvaterra}, {Turolla}, {Watson}, {Wilms}, \&
  {Wolter}}]{Israel2017}
{Israel}, G.~L., {et~al.} 2017{\natexlab{a}}, Science, 355, 817

\bibitem[{{Israel} {et~al.}(2017{\natexlab{b}}){Israel}, {Papitto}, {Esposito},
  {Stella}, {Zampieri}, {Belfiore}, {Rodr{\'\i}guez Castillo}, {De Luca},
  {Tiengo}, {Haberl}, {Greiner}, {Salvaterra}, {Sandrelli}, \&
  {Lisini}}]{P13Israel2017}
---. 2017{\natexlab{b}}, \mnras, 466, L48

\bibitem[{{Jackson} {et~al.}(2005){Jackson}, {Scargle}, {Barnes}, {Arabhi},
  {Alt}, {Gioumousis}, {Gwin}, {San}, {Tan}, \& {Tsai}}]{Jackson2005}
{Jackson}, B., {et~al.} 2005, IEEE Signal Processing Letters, 12, 105

\bibitem[{{Jetsu} \& {Pelt}(1996)}]{JP96}
{Jetsu}, L., \& {Pelt}, J. 1996, A\&AS, 118, 587

\bibitem[{{Kaaret} {et~al.}(2017){Kaaret}, {Feng}, \& {Roberts}}]{KFR2017}
{Kaaret}, P., {Feng}, H., \& {Roberts}, T.~P. 2017, \araa, 55, 303

\bibitem[{{Kawashima} {et~al.}(2016){Kawashima}, {Mineshige}, {Ohsuga}, \&
  {Ogawa}}]{KMOO2016}
{Kawashima}, T., {Mineshige}, S., {Ohsuga}, K., \& {Ogawa}, T. 2016, \pasj, 68,
  83

\bibitem[{{King}(2008)}]{King2008}
{King}, A.~R. 2008, \mnras, 385, L113

\bibitem[{{King}(2009)}]{King2009}
---. 2009, \mnras, 393, L41

\bibitem[{{Koliopanos} {et~al.}(2017){Koliopanos}, {Vasilopoulos}, {Godet},
  {Bachetti}, {Webb}, \& {Barret}}]{Koliopanos2017}
{Koliopanos}, F., {Vasilopoulos}, G., {Godet}, O., {Bachetti}, M., {Webb},
  N.~A., \& {Barret}, D. 2017, \aap, 608, A47

\bibitem[{{Kong} {et~al.}(2016){Kong}, {Hu}, {Lin}, {Li}, {Jin}, {Liu}, \&
  {Yen}}]{Kong2016}
{Kong}, A. K.~H., {Hu}, C.-P., {Lin}, L. C.-C., {Li}, K.~L., {Jin}, R., {Liu},
  C.~Y., \& {Yen}, D. C.-C. 2016, \mnras, 461, 4395

\bibitem[{{Liang}(1986)}]{Liang86}
{Liang}, E.~P. 1986, \apj, 304, 682

\bibitem[{{Lin} {et~al.}(2021){Lin}, {Wang}, {Hui}, {Takata}, {Yeung}, {Hu}, \&
  {Kong}}]{Lin2021}
{Lin}, L. C.-C., {Wang}, H.-H., {Hui}, C.~Y., {Takata}, J., {Yeung}, P. K.~H.,
  {Hu}, C.-P., \& {Kong}, A.~K.~H. 2021, \mnras, 503, 4908

\bibitem[{{Liu} {et~al.}(2013){Liu}, {Bregman}, {Bai}, {Justham}, \&
  {Crowther}}]{Liu2013}
{Liu}, J.-F., {Bregman}, J.~N., {Bai}, Y., {Justham}, S., \& {Crowther}, P.
  2013, \nat, 503, 500

\bibitem[{{Lomb}(1976)}]{Lomb76}
{Lomb}, N.~R. 1976, Ap\&SS, 39, 447

\bibitem[{Mardia(1972)}]{Mardia72}
Mardia, K. 1972, Statistics of Directional Data, Probability and Mathematical
  Statistics a Series of Monographs and Textbooks (Academic Press)

\bibitem[{McLachlan {et~al.}(2019)McLachlan, Lee, \& Rathnayake}]{FMM}
McLachlan, G.~J., Lee, S.~X., \& Rathnayake, S.~I. 2019, Annual Review of
  Statistics and Its Application, 6, 355

\bibitem[{{Middleton} {et~al.}(2015){Middleton}, {Heil}, {Pintore}, {Walton},
  \& {Roberts}}]{Middleton2015}
{Middleton}, M.~J., {Heil}, L., {Pintore}, F., {Walton}, D.~J., \& {Roberts},
  T.~P. 2015, \mnras, 447, 3243

\bibitem[{{Middleton} {et~al.}(2014){Middleton}, {Walton}, {Roberts}, \&
  {Heil}}]{Middleton2014}
{Middleton}, M.~J., {Walton}, D.~J., {Roberts}, T.~P., \& {Heil}, L. 2014,
  \mnras, 438, L51

\bibitem[{{Mitsuda} {et~al.}(1984){Mitsuda}, {Inoue}, {Koyama}, {Makishima},
  {Matsuoka}, {Ogawara}, {Shibazaki}, {Suzuki}, {Tanaka}, \&
  {Hirano}}]{Mitsuda84}
{Mitsuda}, K., {et~al.} 1984, \pasj, 36, 741

\bibitem[{{Motch} {et~al.}(2011){Motch}, {Pakull}, {Gris{\'e}}, \&
  {Soria}}]{Motch2011}
{Motch}, C., {Pakull}, M.~W., {Gris{\'e}}, F., \& {Soria}, R. 2011,
  Astronomische Nachrichten, 332, 367

\bibitem[{{Motch} {et~al.}(2014){Motch}, {Pakull}, {Soria}, {Gris{\'e}}, \&
  {Pietrzy{\'n}ski}}]{Motch2014}
{Motch}, C., {Pakull}, M.~W., {Soria}, R., {Gris{\'e}}, F., \&
  {Pietrzy{\'n}ski}, G. 2014, \nat, 514, 198

\bibitem[{{Nasa High Energy Astrophysics Science Archive Research Center
  (Heasarc)}(2014)}]{HEAsoft2014}
{Nasa High Energy Astrophysics Science Archive Research Center (Heasarc)}.
  2014, {HEAsoft: Unified Release of FTOOLS and XANADU}

\bibitem[{{Okajima} {et~al.}(2016){Okajima}, {Soong}, {Balsamo}, {Enoto},
  {Olsen}, {Koenecke}, {Lozipone}, {Kearney}, {Fitzsimmons}, {Numata},
  {Kenyon}, {Arzoumanian}, \& {Gendreau}}]{Okajima2016}
{Okajima}, T., {et~al.} 2016, in Society of Photo-Optical Instrumentation
  Engineers (SPIE) Conference Series, Vol. 9905, Space Telescopes and
  Instrumentation 2016: Ultraviolet to Gamma Ray, ed. J.-W.~A. {den Herder},
  T.~{Takahashi}, \& M.~{Bautz}, 99054X

\bibitem[{{Pannuti} {et~al.}(2011){Pannuti}, {Schlegel}, {Filipovi{\'c}},
  {Payne}, {Petre}, {Harrus}, {Staggs}, \& {Lacey}}]{Pannuti2011}
{Pannuti}, T.~G., {Schlegel}, E.~M., {Filipovi{\'c}}, M.~D., {Payne}, J.~L.,
  {Petre}, R., {Harrus}, I.~M., {Staggs}, W.~D., \& {Lacey}, C.~K. 2011, \aj,
  142, 20

\bibitem[{{Pinto} {et~al.}(2017){Pinto}, {Alston}, {Soria}, {Middleton},
  {Walton}, {Sutton}, {Fabian}, {Earnshaw}, {Urquhart}, {Kara}, \&
  {Roberts}}]{Pinto2017}
{Pinto}, C., {et~al.} 2017, \mnras, 468, 2865

\bibitem[{{Qiu} {et~al.}(2015){Qiu}, {Liu}, {Guo}, \& {Wang}}]{Qiu2015}
{Qiu}, Y., {Liu}, J., {Guo}, J., \& {Wang}, J. 2015, \apjl, 809, L28

\bibitem[{{Remillard} {et~al.}(2021){Remillard}, {Loewenstein}, {Steiner},
  {Prigozhin}, {LaMarr}, {Enoto}, {Gendreau}, {Arzoumanian}, {Markwardt},
  {Basak}, {Stevens}, {Ray}, {Altamirano}, \& {Buisson}}]{Remillard2021}
{Remillard}, R.~A., {et~al.} 2021, arXiv e-prints, arXiv:2105.09901

\bibitem[{{Rodr{\'\i}guez Castillo} {et~al.}(2020{\natexlab{a}}){Rodr{\'\i}guez
  Castillo}, {Israel}, {Belfiore}, {Bernardini}, {Esposito}, {Pintore}, {De
  Luca}, {Papitto}, {Stella}, {Tiengo}, {Zampieri}, {Bachetti}, {Brightman},
  {Casella}, {D'Agostino}, {Dall'Osso}, {Earnshaw}, {F{\"u}rst}, {Haberl},
  {Harrison}, {Mapelli}, {Marelli}, {Middleton}, {Pinto}, {Roberts},
  {Salvaterra}, {Turolla}, {Walton}, \& {Wolter}}]{Rodriguez2020}
{Rodr{\'\i}guez Castillo}, G.~A., {et~al.} 2020{\natexlab{a}}, \apj, 895, 60

\bibitem[{{Rodr{\'\i}guez Castillo} {et~al.}(2020{\natexlab{b}}){Rodr{\'\i}guez
  Castillo}, {Israel}, {Belfiore}, {Bernardini}, {Esposito}, {Pintore}, {De
  Luca}, {Papitto}, {Stella}, {Tiengo}, {Zampieri}, {Bachetti}, {Brightman},
  {Casella}, {D'Agostino}, {Dall'Osso}, {Earnshaw}, {F{\"u}rst}, {Haberl},
  {Harrison}, {Mapelli}, {Marelli}, {Middleton}, {Pinto}, {Roberts},
  {Salvaterra}, {Turolla}, {Walton}, \& {Wolter}}]{RCastillo2020}
---. 2020{\natexlab{b}}, \apj, 895, 60

\bibitem[{{Scargle}(1982)}]{Scargle82}
{Scargle}, J.~D. 1982, ApJ, 263, 835

\bibitem[{{Sch{\"o}nherr} {et~al.}(2007){Sch{\"o}nherr}, {Wilms}, {Kretschmar},
  {Kreykenbohm}, {Santangelo}, {Rothschild}, {Coburn}, \&
  {Staubert}}]{Schonherr2007}
{Sch{\"o}nherr}, G., {Wilms}, J., {Kretschmar}, P., {Kreykenbohm}, I.,
  {Santangelo}, A., {Rothschild}, R.~E., {Coburn}, W., \& {Staubert}, R. 2007,
  \aap, 472, 353

\bibitem[{{Schulz} \& {Mudelsee}(2002)}]{SM2002}
{Schulz}, M., \& {Mudelsee}, M. 2002, Computers and Geosciences, 28, 421

\bibitem[{{Schwarm} {et~al.}(2017{\natexlab{a}}){Schwarm}, {Sch{\"o}nherr},
  {Falkner}, {Pottschmidt}, {Wolff}, {Becker}, {Sokolova-Lapa}, {Klochkov},
  {Ferrigno}, {F{\"u}rst}, {Hemphill}, {Marcu-Cheatham}, {Dauser}, \&
  {Wilms}}]{Schwarm2017I}
{Schwarm}, F.~W., {et~al.} 2017{\natexlab{a}}, \aap, 597, A3

\bibitem[{{Schwarm} {et~al.}(2017{\natexlab{b}}){Schwarm}, {Ballhausen},
  {Falkner}, {Sch{\"o}nherr}, {Pottschmidt}, {Wolff}, {Becker}, {F{\"u}rst},
  {Marcu-Cheatham}, {Hemphill}, {Sokolova-Lapa}, {Dauser}, {Klochkov},
  {Ferrigno}, \& {Wilms}}]{Schwarm2017II}
---. 2017{\natexlab{b}}, \aap, 601, A99

\bibitem[{{Scrucca} {et~al.}(2016){Scrucca}, {Fop}, Brendan, \&
  E.}]{Scrucca2016}
{Scrucca}, L., {Fop}, M., Brendan, M.~T., \& E., R.~A. 2016, The {R} Journal,
  8, 205

\bibitem[{{Staubert} {et~al.}(2014){Staubert}, {Klochkov}, {Wilms}, {Postnov},
  {Shakura}, {Rothschild}, {F{\"u}rst}, \& {Harrison}}]{Staubert2014}
{Staubert}, R., {Klochkov}, D., {Wilms}, J., {Postnov}, K., {Shakura}, N.~I.,
  {Rothschild}, R.~E., {F{\"u}rst}, F., \& {Harrison}, F.~A. 2014, \aap, 572,
  A119

\bibitem[{{Staubert} {et~al.}(2019){Staubert}, {Tr{\"u}mper}, {Kendziorra},
  {Klochkov}, {Postnov}, {Kretschmar}, {Pottschmidt}, {Haberl}, {Rothschild},
  {Santangelo}, {Wilms}, {Kreykenbohm}, \& {F{\"u}rst}}]{Staubert2019}
{Staubert}, R., {et~al.} 2019, \aap, 622, A61

\bibitem[{{Steiner} {et~al.}(2009){Steiner}, {Narayan}, {McClintock}, \&
  {Ebisawa}}]{Steiner2009}
{Steiner}, J.~F., {Narayan}, R., {McClintock}, J.~E., \& {Ebisawa}, K. 2009,
  \pasp, 121, 1279

\bibitem[{{Strohmayer} \& {Mushotzky}(2003)}]{SM2003}
{Strohmayer}, T.~E., \& {Mushotzky}, R.~F. 2003, \apjl, 586, L61

\bibitem[{{Sutton} {et~al.}(2013){Sutton}, {Roberts}, \&
  {Middleton}}]{Sutton2013}
{Sutton}, A.~D., {Roberts}, T.~P., \& {Middleton}, M.~J. 2013, \mnras, 435,
  1758

\bibitem[{{Tiengo} {et~al.}(2013){Tiengo}, {Esposito}, {Mereghetti}, {Turolla},
  {Nobili}, {Gastaldello}, {G{\"o}tz}, {Israel}, {Rea}, {Stella}, {Zane}, \&
  {Bignami}}]{Tiengo2013}
{Tiengo}, A., {et~al.} 2013, \nat, 500, 312

\bibitem[{{Vasco} {et~al.}(2013){Vasco}, {Staubert}, {Klochkov}, {Santangelo},
  {Shakura}, \& {Postnov}}]{Vasco2013}
{Vasco}, D., {Staubert}, R., {Klochkov}, D., {Santangelo}, A., {Shakura}, N.,
  \& {Postnov}, K. 2013, \aap, 550, A111

\bibitem[{{Vasilopoulos} {et~al.}(2020){Vasilopoulos}, {Lander}, {Koliopanos},
  \& {Bailyn}}]{VLKB2020}
{Vasilopoulos}, G., {Lander}, S.~K., {Koliopanos}, F., \& {Bailyn}, C.~D. 2020,
  \mnras, 491, 4949

\bibitem[{{Walton} {et~al.}(2016){Walton}, {F{\"u}rst}, {Bachetti}, {Barret},
  {Brightman}, {Fabian}, {Gehrels}, {Harrison}, {Heida}, {Middleton}, {Rana},
  {Roberts}, {Stern}, {Tao}, \& {Webb}}]{Walton2016}
{Walton}, D.~J., {et~al.} 2016, \apjl, 827, L13

\bibitem[{{Walton} {et~al.}(2018){Walton}, {F{\"u}rst}, {Harrison}, {Stern},
  {Bachetti}, {Barret}, {Brightman}, {Fabian}, {Middleton}, {Ptak}, \&
  {Tao}}]{Walton2018}
---. 2018, \mnras, 473, 4360

\bibitem[{{Whitehurst} \& {King}(1991)}]{WK91}
{Whitehurst}, R., \& {King}, A. 1991, \mnras, 249, 25

\bibitem[{{Wilms} {et~al.}(2000){Wilms}, {Allen}, \& {McCray}}]{WAM2000}
{Wilms}, J., {Allen}, A., \& {McCray}, R. 2000, \apj, 542, 914

\end{thebibliography}

\end{document}